\documentclass[preprint, 3p, authoryear]{elsarticle}
\usepackage{graphicx}
\usepackage{amsmath}
\usepackage{amssymb}
\usepackage{listings}
\usepackage{longtable, lscape}
\begin{document}

\begin{frontmatter}

\title{On the origin of elemental abundances in the terrestrial planets}

\author[rvt]{S.˜Elser\corref{cor1}}
\ead{selser@physik.uzh.ch}

\author[focal]{M.R.Meyer}
\ead{mmeyer@phys.ethz.ch}

\author[rvt]{B.˜Moore}
\ead{moore@physik.uzh.ch}

\cortext[cor1]{Corresponding author. Phone: +41 44 635 5803}

\address[rvt]{Universit\"at Z\"urich, Winterthurerstrasse 190, CH-8057 Z\"urich, Switzerland}
\address[focal]{ETH Z\"urich, Institut f\"ur Astronomie, Wolfgang-Pauli-Strasse 27, CH-8093 Z\"urich, Switzerland}

\begin{abstract}

The abundances of elements in the Earth and the terrestrial planets provide the initial conditions for life and clues as to the history and formation of the Solar System. We follow the pioneering work of \cite{Bond10} and combine circumstellar disk models, chemical equilibrium calculations and dynamical simulations of planet formation to study the bulk composition of rocky planets. We use condensation sequence calculations to estimate the initial abundance of solids in the circumstellar disk with properties determined from time dependent theoretical models. We combine this with dynamical simulations of planetesimal growth that trace the solids during the planet formation process and which include the effects of gravitational and hydrodynamical mixing. We calculate the elemental abundances in the resulting rocky planets and explore how these vary with the choice of disk model and the initial conditions within the Solar Nebula.

Although certain characteristics of the terrestrial planets in the Solar System could be reproduced, none of our models could reproduce the abundance properties of all the planets.
We found that the choice of the initial planetesimal disk mass and of the disk model has a significant effect on composition gradients. Disk models that give higher pressure and temperature result in larger variations in the bulk chemical compositions of the resulting planets due to inhomogeneities in the element abundance profiles and due to the different source regions of the planets in the dynamical simulations. We observed a trend that massive planets and planets with relatively small semi-major axes are more sensitive to these variations than smaller planets at larger radial distance. Only these large variations in the simulated chemical abundances can potentially explain the diverse bulk composition of the Solar System planets, whereas Mercury's bulk composition can not be reproduced in our approach.

\end{abstract}


\end{frontmatter}

\section{Introduction}

The process of planet formation is far from understood. The established scenario for the formation of the Earth and other terrestrial planets is that most of their masses were built up through the gravitational collisions and interactions of smaller bodies \citep{Chamberlin05,Safranov69,Lissauer93}. These smaller bodies, often called planetesimals, form via the coagulation of dust in the midplane of a circumstellar disk. This transition from dust particles to gravitationally interacting bodies is an important topic of ongoing research \citep{Weidenschilling80, Johansen07, Chiang10}.

Many direct simulations of the formation of rocky planets from planetesimals and protoplanets have been performed \citep{Chambers98,Agnor99,Raymond04,OBrien06,Kokubo06}. One of the goals of these numerical studies is to reproduce the inner part of the Solar System, primarily the number of planets and their mass distribution and orbital parameters. Additional constraints can be the delivery of water rich material onto the planets \citep{Raymond04} or the timing of a Moon forming impact taking place \citep{Touboul07}. Dynamical simulations by \cite{Morishima10} include gas drag and type-I migration and the secular perturbative effects of Jupiter and Saturn. Some of the simulated planetary systems partially satisfy the main constraints given by our Solar System but there are still some issues to resolve. For example, the small mass of Mars is not reproduced in most of the dynamical simulations, although a possible solution to this was recently proposed by \cite{Walsh11}.

Determining the composition of the terrestrial planets is a challenging task. In case of the Earth, samples from the crust can be easily obtained and lava provides samples from the mantle. There are samples from Mars in the form of meteorites and in situ analyses of rocks from spacecrafts and probes on Venus and Mars. Remote sensing and geophysical measurements provide additional data for all planets. Modelling has to include many different processes that are involved in creating the final bulk chemical composition of a terrestrial planet. It reveals that the bulk composition of the planets shows some significant differences in the relative abundance of the major rock forming elements \citep{Morgan80,Kargel93,Lodders97} or their metallic core masses, e.g. \cite{Righter06}. Most meteorites and the rocky planets show a depletion in volatile elements relative to primitive carbonaceous CI meteorites, which resulted from the formation of solids in the protoplanetary disk \citep{Davis06} or from volatile loss during the planet formation process. \cite{Albarede09} pointed out that the volatile depletion of the Earth can be explained by a very hot protoplanetary disk in the inner part of the Solar System, which suppresses the condensation of volatile elements. On the other hand, the Earth is enriched in refractory elements \citep{Palme03,Rubie11}. 

When considering the final elemental compositions of the planets it is also important to consider
the effects of dynamical evolution. The planetesimals that eventually come together to form a terrestrial planet, themselves formed in various locations in the protoplanetary disk. Their bulk composition can depend strongly on their initial location since the formation of solid species in the protoplanetary disk will most likely depend on the local gas composition, temperature and pressure. Gravitational scattering among the planetesimals and protoplanets and radial movement due to interaction with the gas disk results in a radial mixing of the building blocks of the planets and in different bulk chemical compositions of the final planets.

\cite{Bond10} introduced the bulk chemical composition of the Solar System's planets as an additional constraint on the models of their formation. Based on an $\alpha$-disk model by \cite{Hersant01}, they determined the spatial equilibrium composition of solid material in the inner protoplanetary disk with the HSC Chemistry software package. These spatial solid material abundances were mapped onto the initial planetesimals in the dynamical simulations of \cite{OBrien06}. The initial location of each planetesimal that is embodied in the final planets is traced in those simulations. Assuming that the bulk composition of the final proto-planets is the sum of the solids embodied in the initial planetesimals, they estimated the bulk chemical composition of the simulated planets. They found that these are comparable to the bulk chemical composition in the Solar System in the case of most elements, although volatiles are an exception and were often overestimated. Furthermore, \cite{Bond10a} applied their method to extrasolar giant planet systems. They performed dynamical simulations designed to study the formation of rocky planets around those stars with known gas giants detected via radial velocity and known abundances from high resolution stellar spectra. They estimated the bulk chemical composition of the simulated planets and sometimes extreme variations in composition were found. 

In this report, we followed the methodology of \cite{Bond10}. We combined different disk models and the dynamical simulations performed by \cite{Morishima10} to estimate the bulk chemical composition of the simulated planets. \cite{Bond10} were interested in "making the Earth" and reproducing abundances trends, geochemical ratios and oxidation states of the planets. In addition, they studied the late veneer and the loss of volatiles in collisions. In comparison, we focused on the bulk chemical compositions and on variations that might result from different initial conditions in the dynamical simulations and our assumptions in the modeling of the gas disk. In addition, we did not treat the transition from the disk model to the dynamical simulations as a free parameter. Instead, we chose a more self-consistent transition. With those differences, we hope to improve the methodology. In the following, our \textit{disk models} give a time dependent description of the density, temperature and pressure profile of a protoplanetary disk - the physical conditions for the equilibrium chemistry calculations.

The goal of this work consisted of three parts: First, we wished to infer how the distribution of solids in the radial direction in the protoplanetary disk depended on the choice of the disk model. \cite{Bond10} is only based on one model, whereas we explored three simple but widely used disk models and their resulting chemical profiles. This allowed us to estimate how much the bulk composition of planets depends on the conditions in the disk. 

Second, we focused on the dynamical simulations and tried to understand the link between the initial locations of the planetesimals embodied in simulated planets and the planet's final position and mass. A detailed understanding of this link gives deeper insights into planet formation from the dynamical point of view. We compared the cumulated initial distribution of groups of planets to infer if this can be another indicator of trends in the composition of planets. 

Finally, we studied the dependence of the estimated bulk composition of the simulated planets on the disk model and the initial conditions of the dynamical simulations. We were especially interested in the most extreme cases that can be obtained, since such an extreme choice can probably have a strong effect on the results in bulk composition studies. Hence, we could cover a broader range of initial assumptions than \cite{Bond10}. We compared our results with Solar System bulk compositions to see if this modeling can reproduce earlier predictions.

There are uncertainties in the modelling of the protoplanetary disk and assumptions are made in the determination of the equilibrium composition. The transition from the disk model to the disk of solids used in the dynamical simulations is performed in a simple way. The dynamical simulations represent different sets of initial conditions and their effect on the formation process is only poorly understood. The biggest challenge in this work was to deal with those issues and to investigate how the final results were affected by them. We also keep in mind that fundamental processes like disequilibrium condensation and core-mantle differentiation are not taken into account although they might play a crucial role in determining the bulk chemical composition of a planet.

This report is structured as follows: In section \ref{sec:Methodology}, we review the disk models that are used. We briefly present the tool to calculate the equilibrium abundance in the disk and the dynamical simulations. In section \ref{sec:Results}, the bulk chemical compositions of the planets are shown and the differences that arise from the different disk models and initial conditions. In section \ref{sec:Implications}, the caveats and implications of our approach are summarized. Finally, the most important conclusions are given in section \ref{sec:Conclusions}.

\section{Methodology}
\label{sec:Methodology}

Our method consists of a two-step approach first treating the condensation of solids in an evolving circumstellar disk and then following the reshuffling of those materials dynamically through the planet formation process. The evolution of the surface density, pressure and temperature profiles in the disk are given by a disk model. Based on these profiles, a model that describes the chemical processes in the disk gives the abundance of solid species in the disk midplane in space and time. The disk can evolve until it satisfies a selected criterion based on the initial surface density in the dynamical simulations. At this transition time, the distribution of solids in the disk midplane based on the disk model gives the bulk composition of the initial planetesimals in the dynamical simulations. The following dynamical interactions of the particles and their mergers link the bulk composition of the initial planetesimals to the bulk chemical composition of the final planets. 

\subsection{Disk models}

Observations reveal that protoplanetary disks are not static structures. The observed lifetimes are several millions of years over which period mass and angular momentum are transported in the radial direction. The evolution of the disk's surface density $\Sigma$ follows from the conservation of angular momentum and mass and can be expressed in the evolution equation, which is given in the case of a thin disk by \cite{Lynden-Bell74}:
\begin{equation}
\frac{\partial\Sigma}{\partial t}=\frac{3}{r}\frac{\partial}{\partial r}\left(r^{1/2}\frac{\partial}{\partial r}\left(\Sigma\nu r^{1/2}\right)\right).
\label{eq:evolution}
\end{equation}
The angular momentum transport is controlled by the viscosity $\nu$. In the widely-used $\alpha$-prescription, the viscosity is expressed in the form 
\begin{equation}
\nu=\alpha c_s H=\alpha c_s^2 \Omega^{-1}
\label{eq:nu}
\end{equation} 
where $c_s$ is the local sound speed and $H$ is the local scale height of the disk. They are related by $c_s=\Omega H$, with the Keplerian angular motion
\begin{equation}
\Omega=\sqrt{\frac{G M_{\odot}}{r^3}},
\end{equation}
where $G$ is the gravitational constant and $M_{\odot}$ is the mass of the central star, which in our model we always assume to be a solar mass star.
$\alpha$ is a dimensionless quantity which controls the efficiency of angular momentum transport due to viscosity and was introduced by \cite{Shakura73}. This parametrization reflects the uncertainties about the processes causing viscosity. 
The energy dissipation due to viscosity is one of the main heat sources in the disk, especially close to the central star. At larger radii, the irradiation by the central star becomes dominant.  

The isothermal sound speed is

\begin{equation}
c_s=\sqrt{\frac{k_b T_c}{\mu m_H}},
\label{eq:Tc}
\end{equation}

where $k_b$ is the Boltzmann constant, $\mu$ the mean molecular weight and $m_H$ the mass of a hydrogen atom. We assume that all the energy is dissipated close to the midplane of the disk with $T_c$ being the midplane temperature.

Two and even three-dimensional radiative transfer models of circumstellar disks are powerful tools to explore their physical structure and chemical evolution. Very sophisticated models of circumstellar accretion disks were developed in the last years (see \cite{Dullemond07} for a review).  Here, we restrict ourselves to simple analytic models. 
We explore three different disk models here: Two analytic one-dimensional models and a two-dimensional model. In what follows, we derive the surface density, the temperature and the pressure profiles that result from each model.

\subsubsection{Self similar solution}

The self-similar solution of the evolution equation (\ref{eq:evolution}) derived by \cite{Lynden-Bell74} has the form
\begin{equation}
\Sigma(r,t)=\frac{C}{3\pi \nu }A^{-(5/2-\gamma)/(2-\gamma)}\exp\left(-\frac{\tilde r^{(2-\gamma)}}{A}\right),
\end{equation}
where 
\begin{equation}
A=\frac{t}{t_s}+1,
\end{equation}
\begin{equation}
t_s=\frac{1}{3(2-\gamma)^2}\frac{r_s^2}{\nu},
\end{equation}
with $\tilde r=r/r_s$, where $r_s$ is the scale length. It gives the transition radius at which the disk's exponential cut-off starts. This self-similar solution holds if the viscosity is given by a power law, $\nu(r) \propto (r/r_s)^{\gamma}$. We adopt $\gamma=1$. In the $\alpha$-prescription, (\ref{eq:nu}) and (\ref{eq:Tc}) result in a temperature profile with a fixed power law in r:
\begin{equation}
T_c(r)\propto \tilde r^{-1/2}.
\end{equation}
Then, $\nu(r_s)$ is obtained by (\ref{eq:nu}). $C$ is a constant and can be estimated if the initial disk mass is known. 

In principle, the above equation describes a time-independent midplane temperature. Since the surface density evolves in time, we are interested in a time-dependent temperature profile. Hence, we use the above power-law of $T_c\propto \tilde r^{-1/2}$ and derive a surface density-dependent normalization of the profile. To estimate the temperature at $r_s$, we follow \cite{Armitage}, assuming that the disk surface emits like a black body and the vertical energy flux equals the dissipation rate at the disk midplane and is constant in height. The temperature of the disk at $r_s$ is given by
\begin{equation}
T_c^4(r_s,t)=\frac{27 G M_{\odot}}{64\sigma}\kappa \nu(r_s)\Sigma(r_s,t)^2 r_s^{-3},
\end{equation}
where $\kappa$ is the opacity and $\sigma$ is the Stefan-Boltzmann constant. The normalization of the temperature profile becomes time dependent. This approach is not fully self-consistent but it provides a simple model. Finally, the pressure $P$ is given by the ideal gas equation of state by:
\begin{equation}
P(r,t)=\frac{\Sigma}{2}\sqrt{\frac{GM_{\odot}}{r^3}\frac{k_BT_c(r,t)}{m_H \mu  }}.
\label{eq:pressure}
\end{equation}

\subsubsection{Chambers model}

A problem that occurs in the self-similar solution model is that the viscosity is constant in time even when the disk cools and the opacity is constant with radius even very close to the star, where all grains will sublimate. \cite{Chambers09} describes another analytic model which takes into account heating by viscous accretion and radiation from the central star. Three different regimes are introduced: in the innermost disk, where dust grains are sublimated, the opacity follows a power law. Then, for the intermediate part of the disk, a constant opacity is assumed. Viscous heating is the dominant heat source in these regimes. In the third regime, the outer disk, stellar irradiation is the most dominant heat source.

In order to obtain a first approximation, in most parts of the disk we use a constant opacity $\kappa_0$. Although more complex, piecewise continuous opacity laws are used widely (\cite{Ruden91}, \cite{Stepinski98}), there are still uncertainties in the details.

In the intermediate regime, where only viscous heating takes place and the opacity is constant, the disk surface density and temperature are given as a function of time and radius as follows:
\begin{equation}
\Sigma(r,t)=\Sigma_{\rm vis}\left(\frac{r}{s_0}\right)^{-3/5}\left(1+\frac{t}{\tau_{\rm vis}}\right)^{-57/80},
\end{equation}

\begin{equation}
T(r,t)=T_{\rm vis}\left(\frac{r}{s_0}\right)^{-9/10}\left(1+\frac{t}{\tau_{\rm vis}}\right)^{-19/40}.
\end{equation}

$\Sigma_{\rm vis}$ and $T_{\rm vis}$ are the initial surface density and temperature at the outer disk radius. They are given by:
\begin{equation}
\Sigma_{\rm vis}=\frac{7M_{\rm disk}(0)}{10\pi s_0^2},
\end{equation}

\begin{equation}
T_{\rm vis}=\left(\frac{27}{64}\frac{\kappa \alpha \gamma k_B}{\sigma \mu m_H}\right)^{1/3}\Sigma_{\rm vis}^{2/3}\Omega_0^{1/3},
\end{equation}

where $\gamma\simeq 1.4$ is the adiabatic index, $\Omega_0=\sqrt{GM_{\odot}/s_0^3}$ is the Keplerian velocity at the initial disk radius and
$s_0$ is the initial outer radius of the disk, which is the scale length of this model. 

The viscous time scale $\tau_{\rm vis}$ is given by
\begin{equation}
\tau_{\rm vis}=\frac{1}{16\pi}\frac{\mu m_H}{\alpha \gamma k_B}\frac{\Omega_0 M_0}{\Sigma_{\rm vis}T_{\rm vis}}.
\end{equation}

As the surface density of a disk decreases with time, the stellar irradiation will become a dominant heating source and an initially fully viscously heated disk will start to become mostly heated by irradiation in the outer regions, beyond a radius $r_t$.  Then,
at $r>r_t$,
\begin{equation}
\Sigma(r,t)=\Sigma_{\rm rad}\left(\frac{r}{s_0}\right)^{-15/14}\left(1+\frac{t}{\tau_{\rm vis}}\right)^{-19/16},
\end{equation}
where
\begin{equation}
\Sigma_{\rm rad}=\Sigma_{\rm vis}\left(\frac{T_{\rm vis}}{T_{\rm rad}}\right),
\end{equation}
with
\begin{equation}
T_{\rm rad}=\left(\frac{4}{7}\right)^{1/4}
\left(\frac{T_\odot k_B R_\odot}{GM_\odot\mu m_H}\right)^{1/7}\left(\frac{R_\odot}{s_0}\right)^{3/7}T_\odot.
\end{equation}
Moreover,
\begin{equation}
T(r,t)=T_{\rm rad} \left(\frac{r}{s_0}\right)^{-3/7}.
\end{equation}
The radius $r_t$ can be expressed as
\begin{equation}
r_t(t)=s_0\left(\frac{\Sigma_{\rm rad}}{\Sigma_{\rm vis}}\right)^{70/33}\left(1+\frac{t}{\tau_{\rm vis}}\right)^{-133/132},
\end{equation}
and this transition radius $r_t$ moves inwards with time. These equations hold if the initial outer edge of the disk is always in the viscous regime, $s_0<r_t$, which means that irradiation can be initially neglected at the outer disk edge. This is a suitable assumption in our case.

In the regime very close to the star, \cite{Chambers09} assume that the opacity $\kappa$ depends on the temperature
\begin{equation}
\kappa=\kappa_0\left(\frac{T}{T_e}\right)^n,
\end{equation}
where $T_e$ is a transition temperature which defines the border between the inner most region and the intermediate regime. \cite{Stepinski98} recommended $T_e=1380\,$K and $n=-14$. 
The surface density and temperature change with time as follows:
\begin{equation}
\Sigma(r,t)=\Sigma_{\rm evap}\left(\frac{r}{s_0}\right)^{-24/19}\left(1+\frac{t}{\tau_{\rm vis}}\right)^{-17/16},
\end{equation}
with
\begin{equation}
\Sigma_{\rm evap}=\Sigma_{\rm vis}\left(\frac{T_{\rm vis}}{T_e}\right)^{14/19},
\end{equation}
and
\begin{equation}
T(r,t)=T_{\rm vis}^{5/19}T_e^{14/19}\left(\frac{r}{s_0}\right)^{-9/38}\left(1+\frac{t}{\tau_{\rm vis}}\right)^{-1/8}.
\end{equation}

The transition radius between the inner and the outer viscous regimes has the form
\begin{equation}
r_e(t)=s_0\left(\frac{\Sigma_{\rm evap}}{\Sigma_{\rm vis}}\right)^{95/63}\left(1+\frac{t}{\tau_{\rm vis}}\right)^{-19/36}
\end{equation}
and moves inwards with time.

\subsubsection{Two-dimensional model}

The analytic models are given by radius-dependent quantities and basically all quantities in the vertical direction are represented by their central or averaged values. Following \cite{Papaloizou99}, \cite{Hersant01} and \cite{Hure00}, we estimate the pressure and temperature profiles of the disk in a more sophisticated way in a two-dimensional model: in radial direction $r$ and in vertical direction $z$. For each radius, a set of differential equations describes the vertical structure of the disk, where we ignore self-gravity. The structure is described \citep{Armitage} by hydrostatic equilibrium 
\begin{equation}
\frac{1}{\rho}\frac{dP}{dz}=-\Omega^2 z,
\label{eq:hydrostatic}
\end{equation}
by the vertical variation of the flux $F(z)$
\begin{equation}
\frac{dF}{dz}=\frac{9}{4}\rho(z)\nu(z)\Omega^2,
\label{eq:flux}
\end{equation}
and by the radiative heat flux in the diffusion approximation for radiative transfer,
\begin{equation}
\frac{dT}{dz}=-\frac{3\kappa\rho(z)}{16\sigma T(z)^3}F(z).
\end{equation}

 In addition, there is an equation of state 
\begin{equation}
P(z)=\frac{\rho(z)k_B T}{\mu m_H}+\frac{4\sigma}{3c}T(z)^4.
\label{eq:EoS1}
\end{equation}
The second term is the radiation pressure which is usually very small and we ignore it. Turbulent pressure can be ignored if $\alpha\ll 1$, \cite{Hure00}.

The opacity $\kappa$ is in general a function of $P$ and $T$. In this model we use an opacity model presented by \cite{Ruden91}, which is a piecewise, continuous function of temperature. More sophisticated models add an additional atmosphere-like optically thin layer above the optically thick disk \citep{Hure00}, which we neglect. 

\cite{Papaloizou99} used the above set of equations to estimate the vertical structure of protoplanetary disks. The pressure $P(H)$ is, with a boundary optical depth of $\tau(H)=2/3$, given by
\begin{equation}
P(H)=\frac{2H\Omega^2}{3\kappa}.
\end{equation} 
\cite{Papaloizou99} pointed out that the results do not depend on the exact choice of $\tau(H)$ if $\tau(H)\ll 1$. 
At least the midplane pressure does not depend sensitively on this value, although this need not be the case for values at other scale-heights. 

The accretion rate $\dot M_{\rm acc}(t)$ in the disk decreases in time according to (\cite{Drouart99}),
\begin{equation}
\dot{M}_{\rm acc}(t)=\dot M_{\rm acc}(0)\left(1+\frac{t}{t_0}\right)^{-3/2},
\label{eq:MaccHersant}
\end{equation}
where
\begin{equation}
t_0=\left[\frac{R_D^2}{3\nu(R_D)}\right]_{t=0}.
\end{equation}
$R_D$ is the time dependent outer radius of the disk. The initial accretion rate is assumed to be $\dot M_{\rm acc}(0)= 5\times10^{-6}\,M_{\odot}$yr$^{-1}$.

Each annulus of the disk can be calculated independently. To solve the set of equations (\ref{eq:hydrostatic} - \ref{eq:EoS1}), initially, a scale height $H$ is guessed. Calculations start at the top of the disk located at height $H$, where the boundary conditions are given by the optical depth $\tau(H)=2/3$, and
\begin{equation}
F(H)=\frac{3}{8\pi}\dot{M}_{\rm acc}(t)\Omega^2.
\end{equation}
In the midplane, the energy flux along the vertical direction must be $F(0)=0$ due to disk symmetry. If the estimated flux $F_{\rm est}(0)$ in the first calculation is not sufficiently small, that is $\left|F_{\rm est}(0)/F(H)\right|<10^{-4}$, a new $H$ is chosen, estimated through a Newton-Raphson method.
Given the initial disk radius $R_D$, $\alpha$-parameter and the initial accretion rate, we can thus calculate density, temperature and pressure profiles at any time $t$.
\cite{Hure01} reports of a $30\%$ maximum deviation on the midplane temperature comparing one-dimensional and two-dimensional models.

\subsection{Chemistry}

To estimate the chemical composition of solid material in the protoplanetary disk midplane, equilibrium condensation of the gas is assumed. In a closed system of elements, condensation calculation models describe the equilibrium distribution of the elements between coexisting phases. For decades, condensation calculations have been used to describe chemistry in astrophysical problems, for example the formation of the Earth \citep{Latimer50}. Equilibrium condensation provides a basic framework to predict and interpret bulk chemical composition of meteorites and planets \citep{Ebel06,Davis06,Righter06}. Hence, condensation calculations are a good starting point for our study. Following previous studies \citep{Pasek05, Bond10}, we use the commercial HSC Chemistry software package to perform the equilibrium condensation calculations. The equilibrium chemical composition of a system of elements is found by iteratively minimizing it's Gibbs free energy, which is done in the GIBBS equilibrium solver. The algorithm is further described in \cite{White59}. Over one hundred different gaseous and solid species are taken into account, see \cite{Bond10}. The initial parameters for the HSC Chemistry equilibrium composition calculation module are the amount of each pure gaseous element in $kmol$. It is assumed that no other species are present initially. In addition, the solids that are present when equilibrium is reached are also specified. Starting from a gas of solar composition \citep{Asplund05}, see table \ref{tab:element}, and under the assumption of chemical equilibrium, we calculate which distribution of compounds is most thermodynamically stable at midplane pressure $P(r,t)$ and temperature $T(r,t)$ given by the disk models. We do the calculations in a radial range of $r=[0.25\,\rm AU,4\,\rm AU]$ with a step size of $\Delta r=0.05\,$AU.

\begin{table}
\centering
\begin{tabular}{ r  l  r  l }
\hline
Element&Abundance&Element&Abundance\\
\hline
\hline
H&12.00&Si&7.51 \\
He&10.93&P&5.36\\
C&8.39&S&7.14\\
N&7.78&Ca&6.31\\
O&8.66&Ti&4.90\\
Na&6.17&Cr&5.64\\
Mg&7.53&Fe&7.45\\
Al&6.37&Ni&6.23\\
\hline
\end{tabular}
\caption{Relative element abundances in the present-day solar photosphere from \cite{Asplund05}. The abundance is given as the exponent to the base 10. Hence, H has the highest abundance in the solar photosphere, its relative abundance is $10^{12}$. If this value is used as input for the HSC Chemistry module, it can be interpreted as the number of moles of H, whereas the normalization is arbitrary and can be neglected.}
\label{tab:element}
\end{table}

The output is the amount of each compound in $kmol$. We are interested in the relative abundances of the elements in the condensates. Therefore, we weight the abundance of a certain species with the number of atoms of the element included in that certain molecule and its atom weight. The relative abundance is expressed as wt.$\%$ which means \textit{weight percent(age)}:
\begin{equation}
\textrm{wt.}\% \textrm{ of } X=\frac{\sum_A m_X X_A n_A(r)}{\sum_A \sum_Y m_Y Y_A n_A(r)} 
\label{eq:weightpercentage}
\end{equation}
where $n_A$ is the total abundance of species $A$ located at radius $r$ and $X_A$ gives the number of $X$-atoms in species $A$. $m_X$ is the atom weight of the element. The numerator is a summation over all solid species that contain the element $X$. The denominator is a summation over the weight of all atoms $Y$ in all species $A$ in the solid phase at radius $r$. 

In the following, we often order the involved elements according to their volatility. A good measure of the volatility is the temperature at a given pressure at which $50\,\%$ of the original element has condensed \citep{Lodders03}. In the case of O (and C), this temperature is lower than any temperature reached in our disk models, since most O condenses in water ice. Therefore, its volatility is measured according to the condensation of the higher temperature condensates. 

\subsection{Dynamical simulations}
\label{sec:DynamicalSimulations}

\cite{Morishima10} carried out 64 simulations which describe the collisional growth of planetesimals and the subsequent long-term evolution and stability of the resulting planetary systems. The simulations explore sensitivity to the initial conditions, including  the initial mass and radial distribution of planetesimals, the timescale for the dissipation of the solar nebula and different orbits of Jupiter and Saturn. The simulations start with 2000 equal-mass particles placed between 0.5 and $4\,$AU embedded in a gas disk. The planetesimal disk has an initial mass $m_{\rm planetesimals}$ of  $5\,m_\oplus$ or $10\,m_\oplus$. Since we use this mass very often in this report, we introduce m5 for the case of a dynamical simulations with $m_{\rm planetesimals}$=$5\,m_\oplus$ or m10 for the case with $10\,m_\oplus$ The surface density $\Sigma$ of solids and of the initial gas disk depends on the radius, $\Sigma \propto r^{-p}$, where $p$ is 1 or 2. The gas disk dissipates uniformly in space and exponentially in time with a gas dissipation time scale $\tau_{\rm gas}=1, 2, 3$ or $5\,$Myr. After time $\tau_{\rm gas}$ from the beginning of the simulation, when a significant part of the gas disk has disappeared, Jupiter and Saturn are introduced on their orbits. Several orbits are tested: CJS are Circular orbits according to the initial conditions used in the Nice model \citep{Tsiganis05}, EJS place Jupiter and Saturn on their current orbits, EEJS are the same as EJS but higher eccentricity for Jupiter, CJSECC are the same as CJS but higher eccentricities for Jupiter and Saturn. Each simulation represents one unique combination of initial conditions. See tables 1 and 2 in \cite{Morishima10} for further details.

The simulations in \cite{Morishima10} were constructed in an attempt to reproduce the global characteristics of the terrestrial planets in the Solar System. The constraints which were used to compare the simulation results with the observations are the spatial mass distribution, the deviation from circular and coplanar orbits and the timing of the Moon-forming impact. In general, $1-5$ planets are formed in the terrestrial region. An observed trade-off is that if a similar radial mass concentration is achieved, the Moon-forming impact occurs too early. The dependence on the initial conditions becomes manifest for example in the fact that small gas dissipation timescales ($\tau=1-2\,$Myr) are needed to avoid a significant depletion of solid material due to migration. 

Other simulations \citep{OBrien06} often consist of initially roughly Mars-mass embryos embedded in a planetesimal disk with no mutual interaction between the planetesimals. The initial mass of the fully interacting particles in the \cite{Morishima10} simulations is smaller than the Moon mass. Hence, Mercury analogs concerning its mass and semi-major axis can form in those simulations in contrast to other studies. Nevertheless, these calculations assume perfect accretion and no mass loss due to giant impacts - such an event is most probably responsible for striping off Mercury's initial crust and mantle (see \cite{Benz07} for a review). 

The Minimum Mass Solar Nebula (Hayashi, 1981) has a gas density profile $\Sigma \propto r^{-p}$ with $p=1.5$. The steady-state solution of (\ref{eq:evolution}) predicts $p=1$. Observations suggest a flatter profile with $p<1$ \citep{Kitamura02, Andrews09}. Therefore, a gas density profile with $p = 2$ is extremely steep in the context of circumstellar disks and we focus on the dynamical simulations with $p=1$. The initial surface density of solids at $1$ AU is $\Sigma_{0,\rm solid}=6.1\,\rm{g/cm^2}$ in the case of m5 or $\Sigma_{0,\rm solid}=12.2\,\rm{g cm^{-2}}$ for m10. The gas density at 1 AU is $\Sigma_{0,\rm gas}=2000\,\rm{g cm^{-2}}$ for both m5 and m10. 

\subsection{Transition from disk models to dynamical simulations}

In order to combine the disk models and the dynamical simulations, we have to choose a suitable transition criterion. Since our goal is to study the evolution of the solid material in the dynamical simulations, we define the solid surface density at 1$\,$AU as the normalization condition. This means that the surface density of solids $\Sigma_{\rm solid}$ predicted by the disk model at 1$\,$AU at transition time $t=t_{\rm trans}$ has to reproduce the initial solid surface density of the dynamical simulations at 1$\,$AU, $\Sigma_{0,\rm solid}$. For example,
\begin{equation}
\Sigma_{\rm solid}(1\,AU,t_{\rm disk})=\Sigma_{0,\rm solid}=6.1\,\rm{g/cm^2},
\end{equation}
in the case of the m5 simulations.
Actually, the disk models give the surface density of gas $\Sigma_{\rm gas}$ and we assume a gas-to-dust ratio of
\begin{equation}
\frac{\Sigma_{\rm gas}(r,t)}{\Sigma_{\rm solid}(r,t)}=100,
\end{equation}
which is roughly the gas-to-dust ratio of the ISM. This value is smaller than the gas-to-dust ratio in \cite{Morishima10};
\begin{equation}
\frac{\Sigma_{0,\rm gas}}{\Sigma_{0,\rm solid}}=329,
\end{equation}
for m5 simulations, and half that for m10. These ratios are rather high and can only be achieved with a very high solid mass loss rate combined with a low gas dissipation rate. Since we model the solids in this disk, the ISM ratio seems to be a better choice.   

The initial total disk mass in the dynamical simulations is 
\begin{equation}
m_{\rm tot}\approx m_{\rm planetesimals} \frac{\Sigma_{0,\rm gas}}{\Sigma_{0,\rm solid}}=1650\,m_{\oplus}\cong0.0055\, M_{\odot},
\end{equation}
for m5 and the m10 simulations. 
Therefore, the initial mass of $0.1\,M_\odot$ in our disk models is a justified assumption since the disk loses mass with time. Disk model masses at $t=t_{\rm trans}$ are shown in table \ref{tab:ttrans}.

A disk model is described by the initial mass, time $t$, the $\alpha$-parameter and the disk scale radius $r_s$. For all models we adopt the following initial constraints: the initial disk mass $M_{\rm disk}(t=0)=0.1\,\rm{M_{\odot}}$ and $\alpha=0.009$,  \cite{Hersant01}.  The models use different opacity laws. By default, the constant opacity is $\kappa_0=3\,\rm{cm^2\,g^{-1}}$. We adopt a radial scale length $r_s=10\,$AU. We are only interested in profiles that reproduce the surface density at 1$\,$AU. The choice of the scale radius has no significant effect on the results: In the self-similar solution model, the effect of the scale radius $r_s$ on the surface density distribution can be compensated by a suitable choice of $t_{\rm trans}$ and $C$. It has a minor effect on temperature and pressure. 

In the Chambers model, the initial disk radius and the time $t$ are free parameters and they are degenerate in a similar way as in the self-similar solution model. In the two-dimensional model, the accretion rate is a function of the initial outer disk radius, time and initial accretion rate (see equation (\ref{eq:MaccHersant})). Since the surface density in this model is controlled by the accretion rate, the disk model is also degenerate in the parameters $R_D$ and $t_{\rm trans}$. Actually, the choice of the scale radius is not completely free: In the case of a very large initial scale radius ($r_s\sim100\,$AU), the initial disk mass might be too small to reproduce the surface density at $1\,$AU for any $t>0$. For $r_s=10\,$AU, there is a $t_{\rm trans}>0$ such that $\Sigma(1\,AU,t_{\rm disk})=\Sigma_{0,\rm solid}$ for all models. The same holds for the choice of the initial accretion rate in the two-dimensional model. If the initial accretion rate is too small, it will never result in a suitable surface density. Table \ref{tab:ttrans} lists the model-dependent transition times which can differ by orders of magnitude.

Note that in \cite{Bond10}, the transition time is left as a free parameter. Hence, a better fit to the observed planetary abundance values is possible, but this approach is not as self-consistent as ours.

\section{Results}
\label{sec:Results}

This section presents the different disk models that are normalized to the initial conditions of the dynamical simulations. The resulting chemical abundance profiles are shown briefly. Then, the source regions of the planets are studied. Combining the chemistry and the dynamics results in the bulk chemical composition of the planets which are studied in detail. Two elements which are very important for the emergence and evolution of life, H and C, are briefly highlighted at the end of the section.

\subsection{Initial disk profiles}
\label{sec:Initialprofiles}

\begin{figure}
\centering
\includegraphics[scale=0.78]{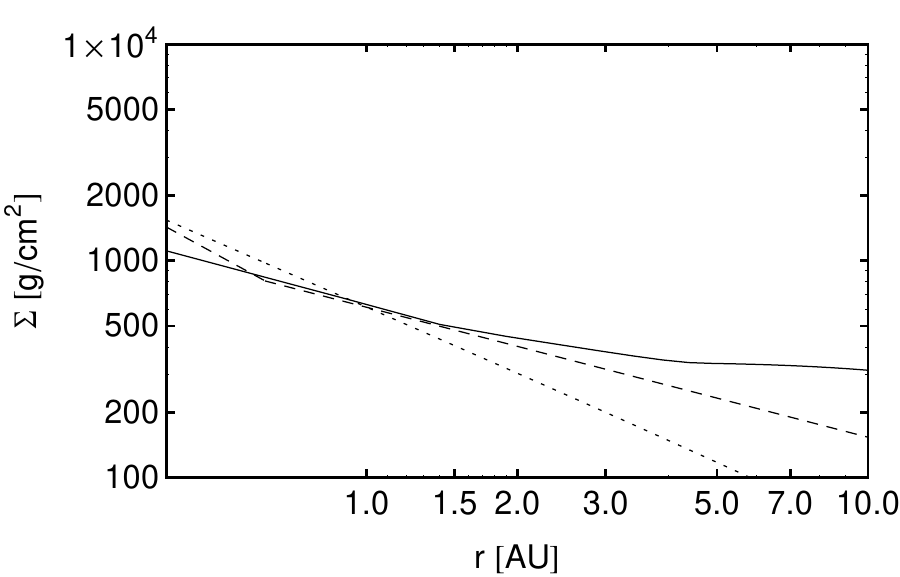}\,\,\,\includegraphics[scale=0.78]{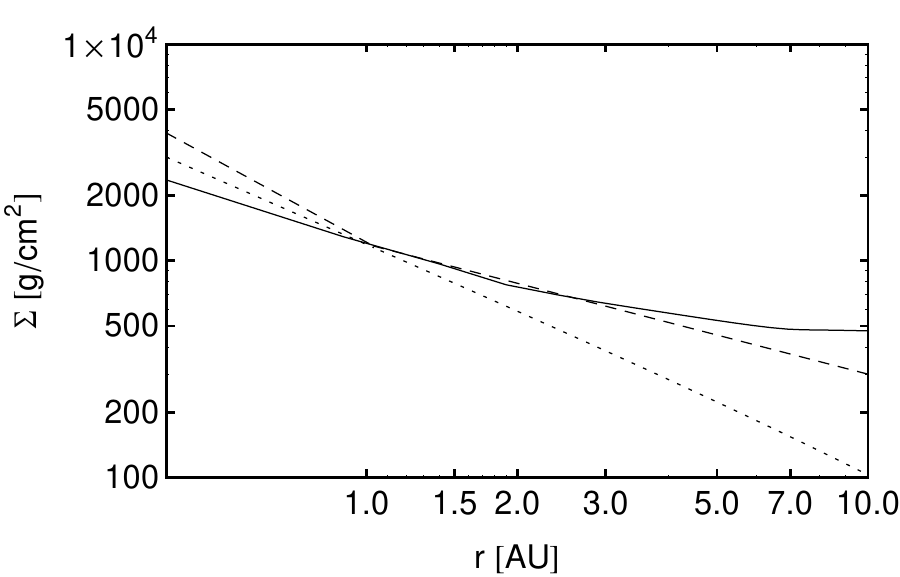}\newline
\includegraphics[scale=0.78]{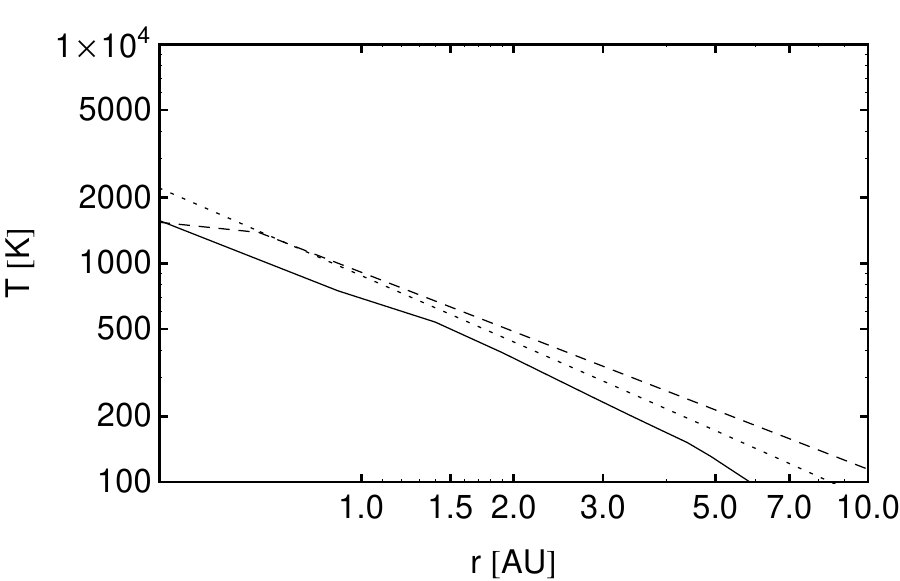}\,\,\,\includegraphics[scale=0.78]{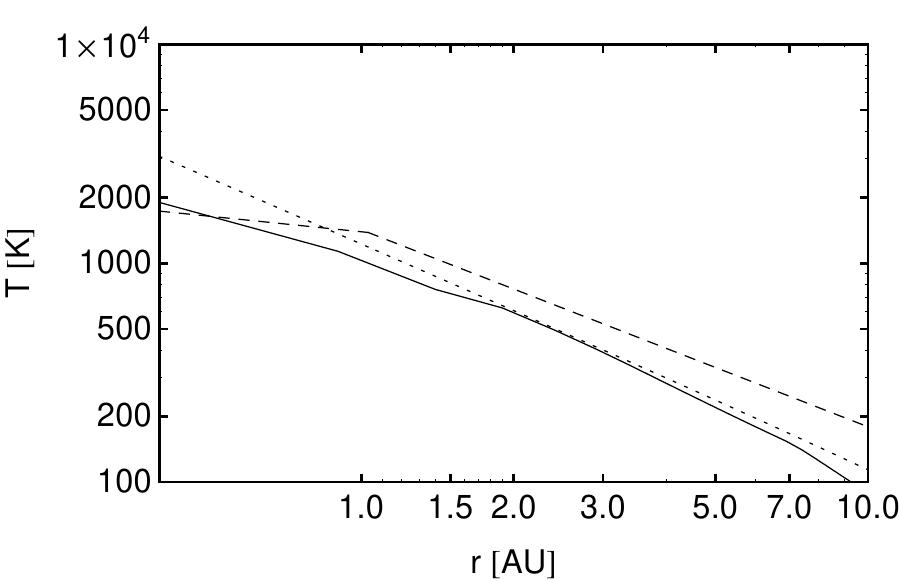}\newline
\includegraphics[scale=0.78]{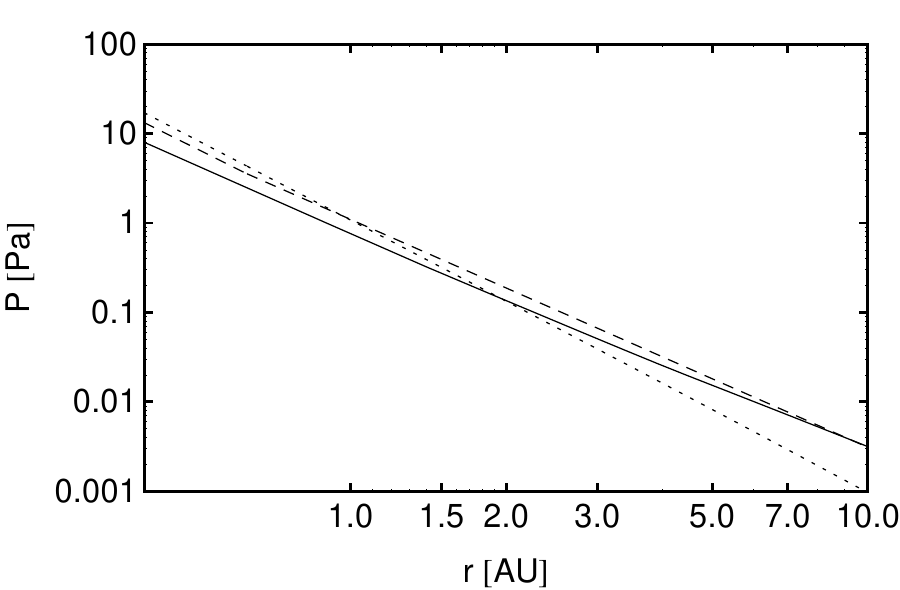}\,\,\,\includegraphics[scale=0.78]{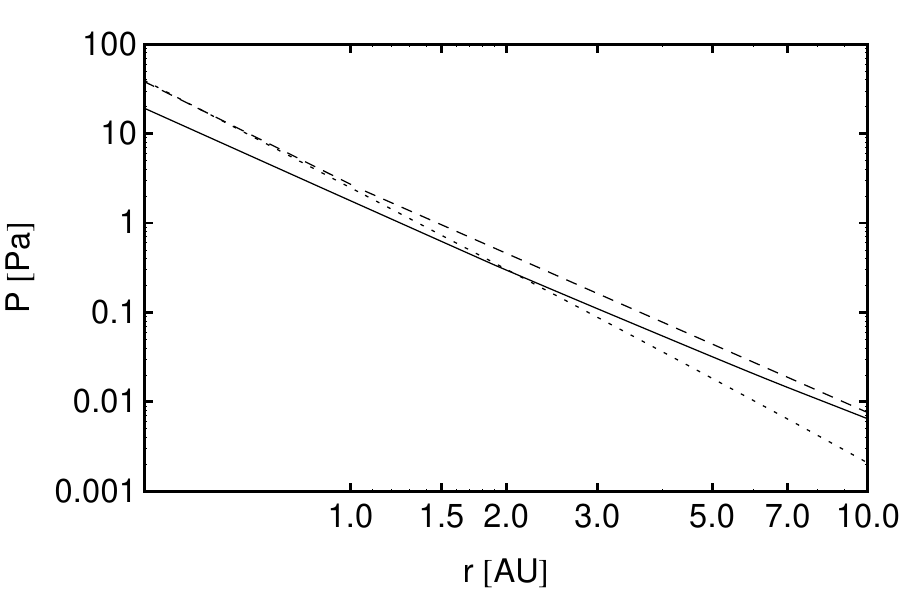}\newline\centering
\caption{From top to bottom: Surface density profiles, temperature profiles and pressure profiles in the midplane, the m5 disk in the left column, the m10 disk in the right column at transition time. Each plot shows the two-dimensional model (solid line), the Chambers model (dashed line) and the self-similar solution model (dotted line). The adopted model parameters are given in table \ref{tab:ttrans}.}
\label{fig:plots}%
\end{figure}

\begin{table}
\centering

\begin{tabular}{ r   l   c   c  c}
\hline
								 disk model	& disk mass & $t_{\rm trans} [10^4\,yr]$&$M_{\rm disk} [10^{-2}m_{\odot}]$& $r_{\rm in}$ [AU]\\
\hline
\hline
self-similar solution &$m_T=5\,m_{\oplus}$ & 3.7 & 3.5  &	0.50															\\
self-similar solution &$m_T=10\,m_{\oplus}$ & 2.1 & 4.4 &	0.75															\\
Chambers              &$m_T=5\,m_{\oplus}$ &5.0 & 5.0 	&	0.30													\\
Chambers              &$m_T=10\,m_{\oplus}$ &1.9 & 6.0 	&	0.50													\\
two-dimensional       &$m_T=5\,m_{\oplus}$ & 25.0 & 1.2 &	0.35															\\
two-dimensional       &$m_T=10\,m_{\oplus}$ & 41.0 & 1.8& 0.65																\\
\hline
\end{tabular}
\caption{The adopted transition time and the resulting disk mass for the different disk models and solid disk masses. $r_{\rm in}$ is the innermost semi-major axis at which solids condense at $t_{\rm trans}$. The model parameters are $\alpha=0.009$ and $r_s=10\,$AU and the initial disk mass is $m_{\rm disk}(0)=0.1\,M_{\odot}$. }
\label{tab:ttrans}
\end{table}

Surface density profiles and the corresponding pressure and temperature profiles at $t_{\rm trans}$ for the different models are plotted in figure \ref{fig:plots}. The slope in the temperature and pressure profiles of the models are very similar. At radial distance $r>1\,$AU, the different opacity models change the slope of the temperature profile. The slope of the dynamical simulation surface density ($\Sigma\propto r^{-1}$) is only reproduced in the self-similar model with the other models  yielding flatter profiles. The high solid surface density of the m10 simulations results in disk models that predict a high temperature and pressure profiles relative to the m5 disk. The Chambers model gives the highest overall temperatures, especially at larger radii. In general, the two-dimensional model gives the lowest temperature. These differences allow us to explore a range of different temperature and pressure profiles in our study.

Comparison with disk profiles obtained in previous studies is difficult since our initial assumptions vary. A brief comparison with e.g. the sophisticated study by \cite{DAlessio98} reveal that our profiles represent a reasonable approximation to more complex models.

\subsection{Elemental abundance profiles}

\begin{figure}
\centering
\includegraphics[scale=0.6]{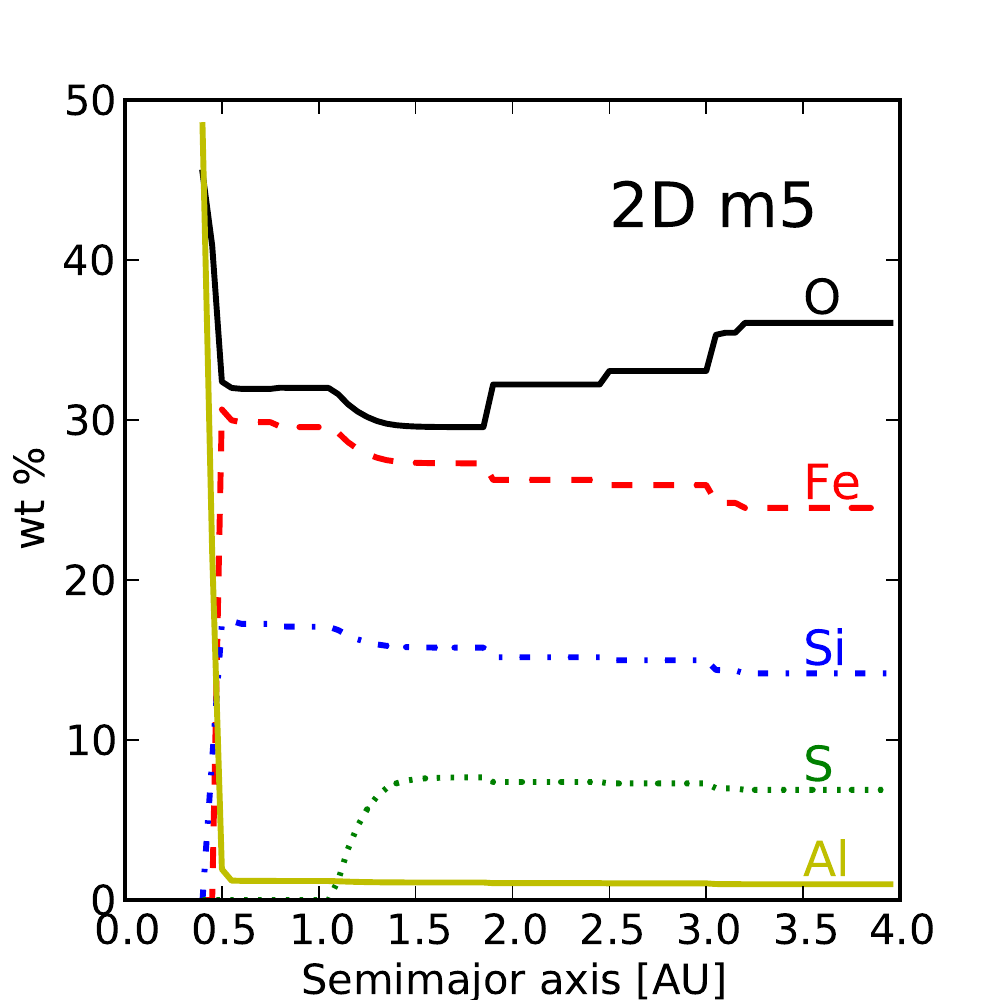}\,\,\,\includegraphics[scale=0.6]{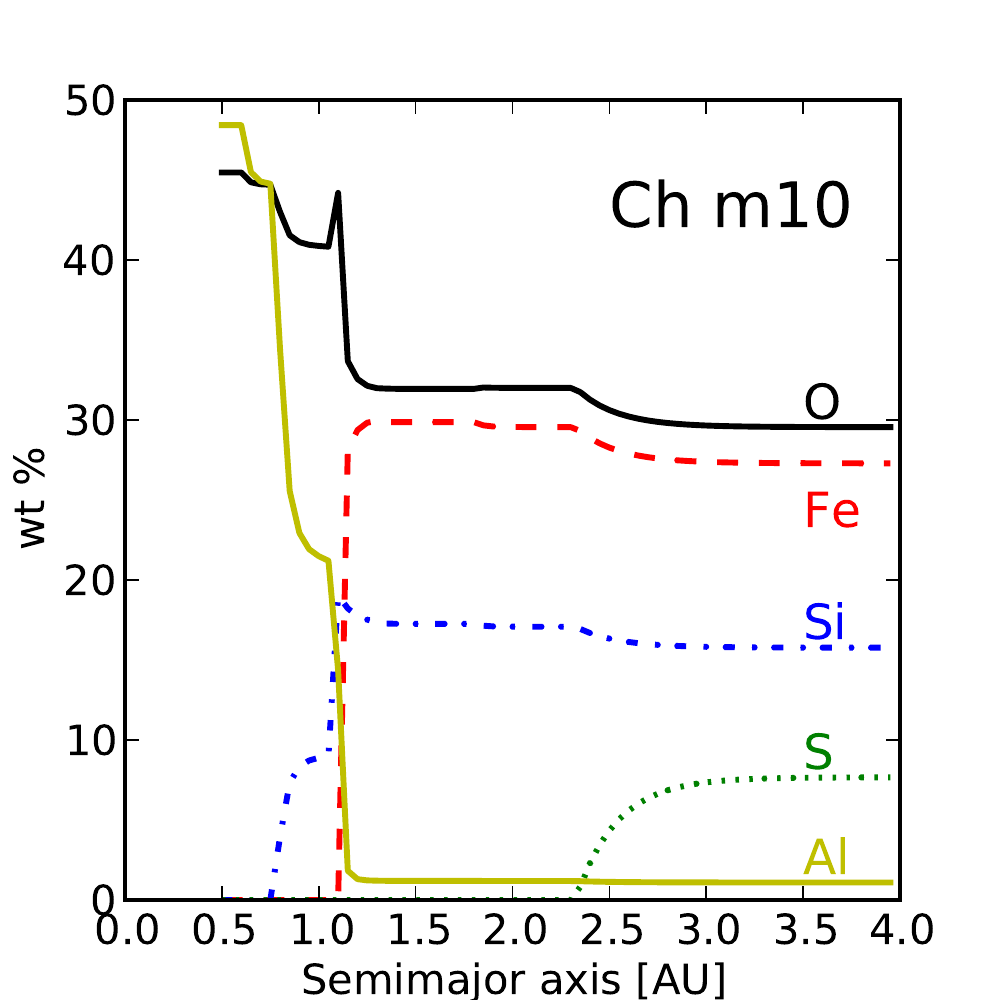}
\caption{The weight percentage chemical abundance of some of the most frequent elements in solid species as a function of distance from the star. The left panel gives the two-dimensional model (2D) in the case m5, and the right panel gives the Chambers disk model (Ch) in the case m10. The elements are from top to bottom at 4 AU: O, Fe, Si, S and Al. For example: in the 2D case, a solid particle condensing at 1$\,$AU is composed by weight of $~30\,\%$ O, $~30\,\%$ Fe, $~20\,\%$ Si and $20\,\%$ is comprised of all other elements.}
\label{fig:chemradius}%
\end{figure}

Figure \ref{fig:chemradius} shows the elemental abundances in solids as a function of radius in the case of the two-dimensional model and a disk of mass m5 and the Chambers disk model in the case m10 as extrema. The two-dimensional model is a good case study to explain some of the main features since this model gives the largest numbers of solids along the disk: at 1.3 AU, troilite (FeS), at 2 AU, fayalite ($\rm{Fe_2SiO_4}$) and around 3 AU iron oxide ($\rm{Fe_3O_4}$) all condense out of the solar nebula. Serpentine ($\rm{Mg_3Si_2O_5(OH)_4}$) and aluminum oxide ($\rm{Al_2O_3}$) follow at 3.1 AU. Since all models have similar temperature profile slopes at $r>1\,$AU but predict different temperature normalizations, the location of condensation shifts to larger radii in the case of colder disks. In the case of the Chambers model in figure \ref{fig:chemradius}, the condensation of FeS is the only feature beyond $\sim1$ AU beside the occurrence of Fe, magnesium silicates and other species around 1 AU. 

We point out that the total amounts of solid material changes completely with semi-major axis. Hence, traces of refractory solids that condense close to the star can dramatically change the percentage weights of the different elements. We see such features in the profile close to the star. If the temperature and pressure are high enough ($T \gtrsim 1340\,$K, $p\gtrsim10^{-5}\,$bar) so that even refractory solids would sublimate, the estimation of the elemental abundance breaks down. Inside this radius $r_{\rm in}$, listed in table \ref{tab:ttrans}, no condensates will form. The self-similar solution model and the two-dimensional model in the m10 disk fail to provide condensates across the whole annulus $a=[0.5\,\rm AU,4\,\rm AU]$ where planetesimals in the dynamical simulations are initially located.

In general, close to the star, only refractory species condense out of the solar nebula. The abundance of more volatile elements increases with distance from the star. Hence, a hot disk model generates a disk with a relatively high abundance of refractory elements. On the other hand, cooler disks provide higher abundances of volatile elements. 

\subsection{Source regions in the N-body simulations}

In the dynamical simulations, the orbits of the planetesimals and protoplanets change due to various mechanisms, such as close encounters or secular resonances. Gas drag damps the eccentricities of the smallest planetesimals and moves them inwards. If they do not grow fast enough and do not merge with other bodies, they do not decouple from the gas. Hence, the planetesimals at small initial semi-major axis fall into the star. On the other hand, larger bodies start to migrate inward due to type I migration as long as the gas disk has a significant surface density. The gas dissipation time scale is of the order of few Myr. As a result, most of the material initially located inside 1 AU is lost in almost every simulation. 

\begin{figure}
\centering
\includegraphics[scale=0.4]{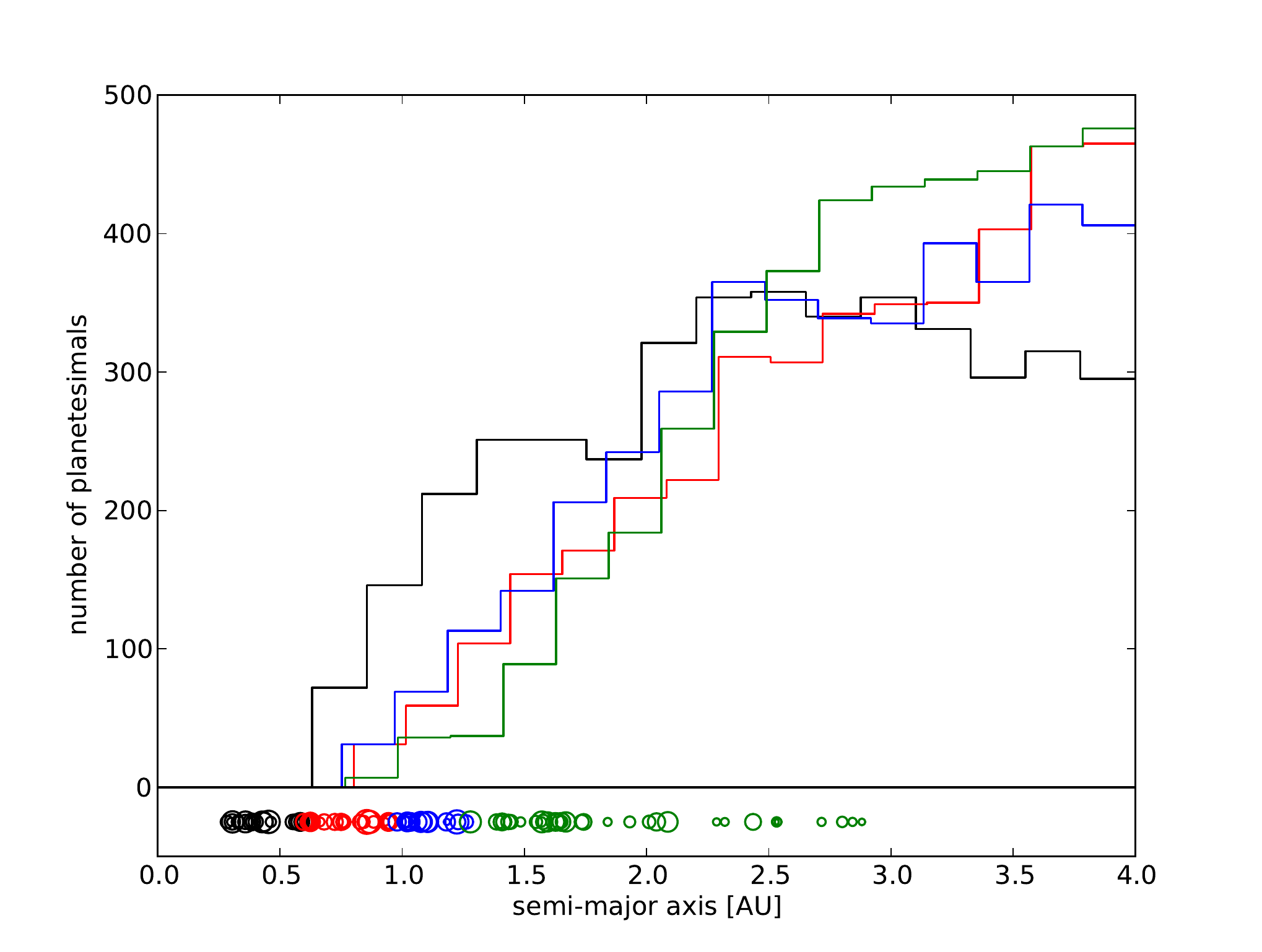}\,\includegraphics[scale=0.4]{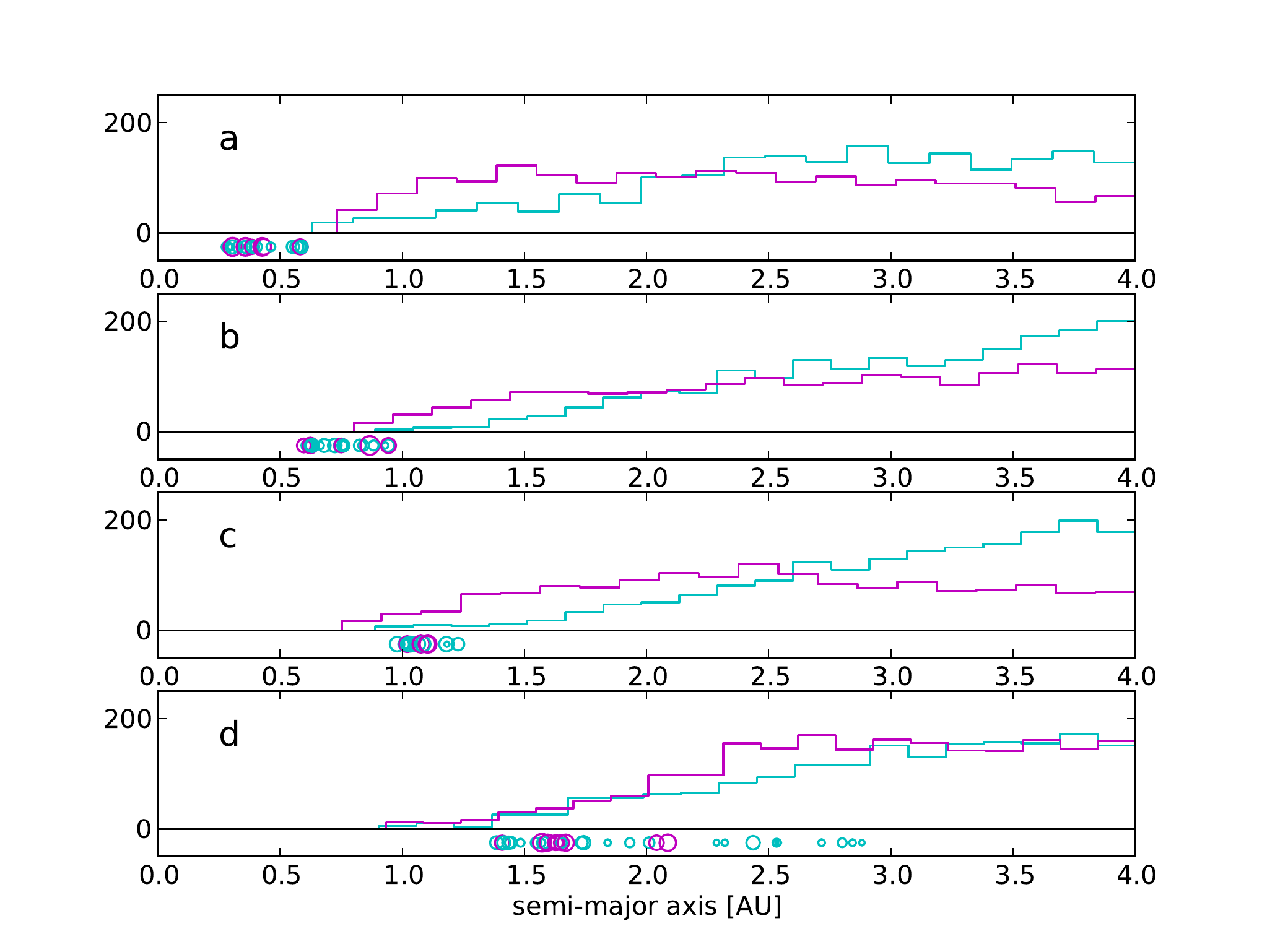}
\caption{Left: The initial distribution of all planetesimals that end up in one of the planets in four orbital groups indicated by different colors ($a$: black, $b$: red, $c$: blue, $d$: green). The lower panel shows the distribution of all planets in radial distance, their diameter represents their masses. Right: The distribution of the planetesimals of the small mass (cyan) and high mass (magenta) sub-groups in the four orbital groups $a$-$d$. The lower panel in each histogram gives the distribution of the orbital group that is considered.}
\label{fig:histo}%
\end{figure}

The region of initial locations of the planetesimals that end up embodied in a final planet is what we call the planet's source region. In order to study the source regions of individual planets, we group planets first according to their semi-major axis. We divide the planets in four groups ($a,b,c$ and $d$, ordered by increasing semi-major axis), each containing the same total mass (and thus the same number of initial planetesimals (see figure \ref{fig:histo})). To study the source region as a function of the planet's final mass in each of these groups, we divide each group $a,b,c$ and $d$ in two sub-groups containing the same number of initial planetesimals: a group of high mass planets and a group of low mass planets. With the statistics  of 4 orbital groups and $4\times2$ mass sub-groups, the dependence of the source region on the planet's semi-major axis and final mass can be studied. A study of the dependence on the simulation initial conditions would be desirable but the number of simulations is too small. A planet's source region is characterised by the cumulative distribution function of planetesimals as a function of radius from which it is built. We compare the source regions of any two groups using the two-sided Kolmogorov-Smirnov-test. The estimated p-value gives the probability that the two distributions of the initial planetesimal locations are drawn from the same parent distribution. If $p$ is small, it is unlikely that the source regions coincide. 

First, we explore how the source regions differ depending on the final orbital radii of the planets.   For each planet in a given orbital group, we gather the initial positions of its building blocks. Figure \ref{fig:histo} gives the histograms of the locations of all the initial planetesimals that end up within  the planets of the four different orbital groups. Planets close to the star have a wide and flat source region (group $a$), they form from material that was also closer to the star initially. Planets at larger distances (groups $b, c$ and $d$) all have steeper and narrower source regions. For these groups, the source regions arise entirely from beyond a radial distance of 1 AU.  This is mainly because almost all material initially within  this location ends up in the star. For example, the black histogram (group $a$) indicates that more than 200 planetesimals ($\approx 5\%$ of the total mass of the group) initially located within $1.0\,$AU end up in the planets in group $a$, whereas according to the green histogram, not more than 10 planetesimals end up in planets of group $d$. KS-tests reveal that the groups do not share the same initial distribution as all p-values are below $10^{-6}$. This means that we should not expect that planets at different semi-major axes have the same source region. The inner most group $a$ stands out most significantly from the others (see figure \ref{fig:histo}). 

Next, we study how the source regions differ depending on the final mass of the planets. The planets in each orbital group are split into two sub groups according to their mass. The KS-test shows that the sub-group containing low mass planets and the sub-group containing high mass planets do not have the same parent distribution in all of the orbital groups. A trend is visible: low mass  planets have a source region that is steeper and narrower, whereas massive planets have a wider and flatter source region (see figure \ref{fig:histo}). 

Often, the planets are located much closer to the star than the average radii of the initial planetesimals from which they were built. This is due to orbital migration. Most of the bulk composition of the planets is contributed by the solids that condense out of the solar nebula at $1\,$AU$ < r < 4\,$AU, except some of the very close-in planets.

\subsection{Combining disk chemistry and N-body simulations}

In the next step, the calculations of the chemical abundance in the disk model and the dynamical simulations are combined. According to our transition criterion, the disk model dependent transition time $t_{\rm disk}$ estimated in section \ref{sec:Initialprofiles} is the time in the disk model evolution at which the chemical composition of solids in the disk represents the composition of the planetesimals at the beginning of the dynamical simulations. The initial radial distribution of the planetesimals that end up in a given planet gives the final bulk chemical composition of this planet. The element abundance profiles in the disk and the diversity in the source regions suggest that different planets have different bulk chemical compositions. All planetesimals from the dynamical simulations that contribute to the mass of any planet originate from beyond the condensation line $r_{\rm in}$ at $t_{\rm trans}$. There the composition of all planets can be fully determined. The element abundance in the planets for each disk model is shown in the online supplementary data.

The simulations by \cite{Morishima10} are designed to reproduce the inner planets of the Solar System, thus, we now compare the element bulk abundance of the simulated planets with the Solar System terrestrial planets.

A normalized abundance $N_X$ of element X relative to Si can be obtained from (\ref{eq:weightpercentage}) via:
\begin{equation}
N_{\rm X,tp}=\frac{\left(\frac{wt.\% \rm{of} X}{wt.\% \rm{of} Si}\right)_{\rm sp}}{\left(\frac{wt.\% \rm{of} X}{wt.\% \rm{of} Si}\right)_{\rm tp}}
\label{eq:Nx}
\end{equation}

where the $wt.\%$ is based on simulated planets (subscript sp) or literature values of terrestrial planets (subscript tp) in our Solar System. The bulk chemical compositions were taken from \cite{Morgan80}, \cite{Kargel93}, and \cite{Lodders97}, see table \ref{tab:abundances}. One should keep in mind that these values are partly based themselves on simulations and that uncertainties lead to absolute errors up to the order of $25\%$ on these abundances \citep{Bond10}. These uncertainties are discussed in the caveats section \ref{sec:caveats}.
\begin{table}
\centering
\begin{tabular}{ l  r  r  r  r   }
\hline
&Mercury$^1$&Venus$^1$&Earth$^2$&Mars$^3$\\
\hline
\hline
Fe, $\%$	&	64.47	&31.17&	32.04&	27.24\\
O, $\%$		&14.44&	30.9	&31.67&	33.75\\
Mg, $\%$	&	6.5	&14.54&	14.8&	14.16\\
Al $\%$		&	1.08	&	1.48&	1.43&	1.21\\
Si, $\%$	&	7.05&	15.82	&14.59	&16.83\\
Ni, $\%$	&	3.66&	1.77	&1.72	&1.58\\
S, $\%$		&0.24	&1.62	&0.89	&2.2\\
Ca,  $\%$	&	1.18&	1.61	&1.6	&1.33\\
C, ppm		&5.1&	468	&44	&2960\\
N, ppm		&0.046&	4.3	&0.59&	180\\
Na, ppm		&	200&	1390&	2450&	5770\\
P, ppm		&	390&	1860&	1200	&1100\\
Ti, ppm	&	630&	850&	800	&650\\
Cr ppm		&	7180	&4060&	3400	&3680\\
\hline
\end{tabular}

\caption{Planetary abundances in wt.$\%$ or weight ppm. References: $^1$\cite{Morgan80}, $^2$\cite{Kargel93}, $^3$\cite{Lodders97}. }
\label{tab:abundances}
\end{table}

Next, we search for a simulation outcome that best reproduces the composition of the inner Solar System planets. We choose a planetary system including four planets out of the m10 simulations and a similar system out of the m5 simulations. Since all planets have to form in the same simulation and the number of simulations is limited, we focus on systems that match the constraints of our Solar System best according to \cite{Morishima10}, keeping in mind that differences in mass and semi-major axis have an effect on the source region. EJS 2-1-5 (short hand for $\tau_{\rm gas}=2\,$Myr, $p=1$, m5 (see section \ref{sec:DynamicalSimulations})) and EJS 3-1-10 are two systems that fulfill many of the constraints, although the masses of the planets are very poorly reproduced: Venus and Earth are too small and the Mars and Mercury analogs are too massive. The source region of close in planets is sensitive to the planets mass and the large Mercury mass in the EJS 2-1-5 simulation might affect the result. The large Mars mass is a general problem that is often observed in dynamical simulations that reproduce the inner Solar System, recently discussed in \cite{Walsh11}. The masses and semi major axis of the chosen simulated planets are shown in table \ref{tab:planets}.

\begin{table}
\centering
\begin{tabular}{ c  c  r  r  r  r}
\hline
& &Mercury&Venus&Earth&Mars\\
\hline
\hline
EJS 2-1-5&mass [$m_{\oplus}$]&0.52	&0.45&	0.79	&0.51\\
& $a$ [AU]&0.31 & 0.59 & 0.94 & 1.39\\
\hline
EJS 3-1-10&mass [$m_{\oplus}$]&0.09&	0.23	&0.34&	0.27\\
&$a$ [AU]&0.67 & 0.89 & 1.26 & 1.54 \\
\hline
Solar System &mass [$m_{\oplus}$] &0.06&	0.81	&1.00&	0.11\\
&$a$ [AU]&0.38 & 0.72 & 1.00 & 1.52 \\
\hline
\end{tabular}

\caption{Parameters of simulated planetary systems that represent the Solar System. $a$ is the semi-major axis of the planet. The Solar System values are taken from \cite{Kargel93}.}
\label{tab:planets}
\end{table}

\begin{figure}
\centering
\includegraphics[scale=0.5]{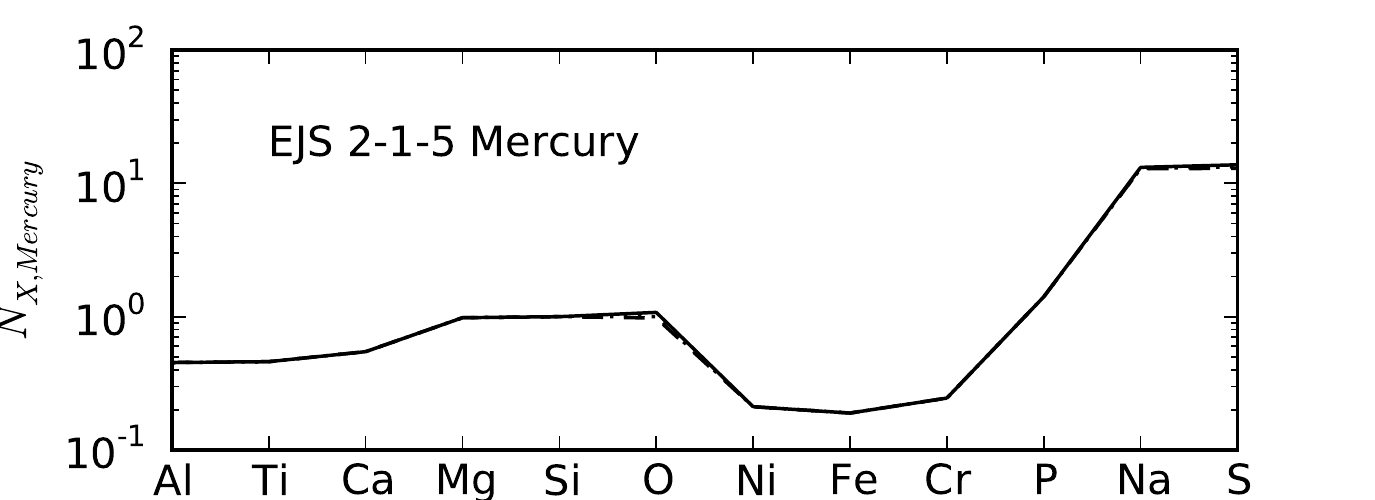}\,\,\,\includegraphics[scale=0.5]{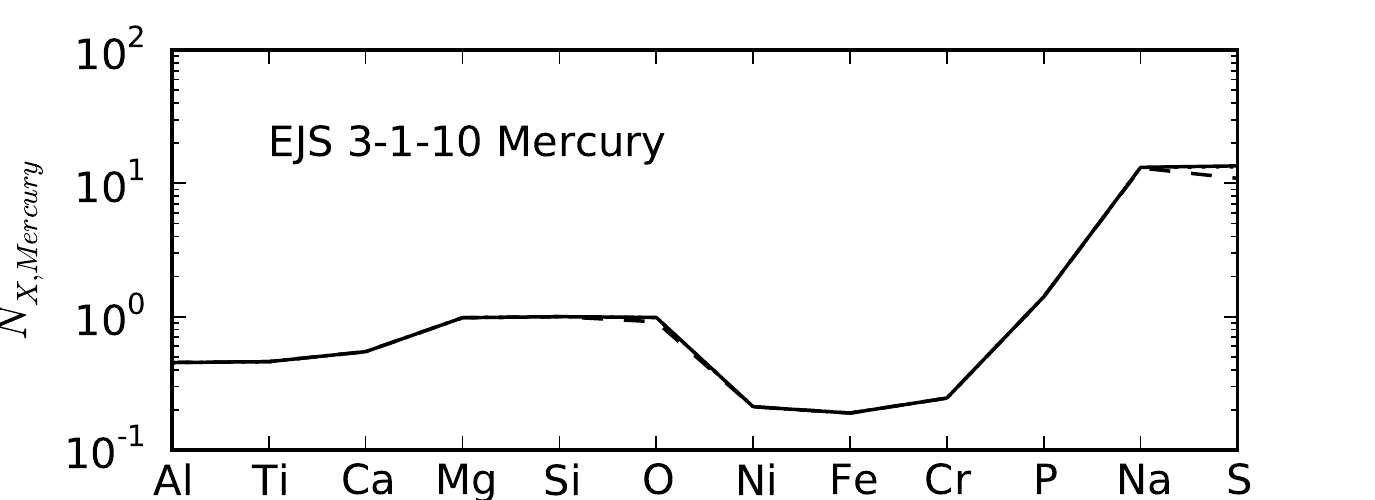}\newline
\includegraphics[scale=0.5]{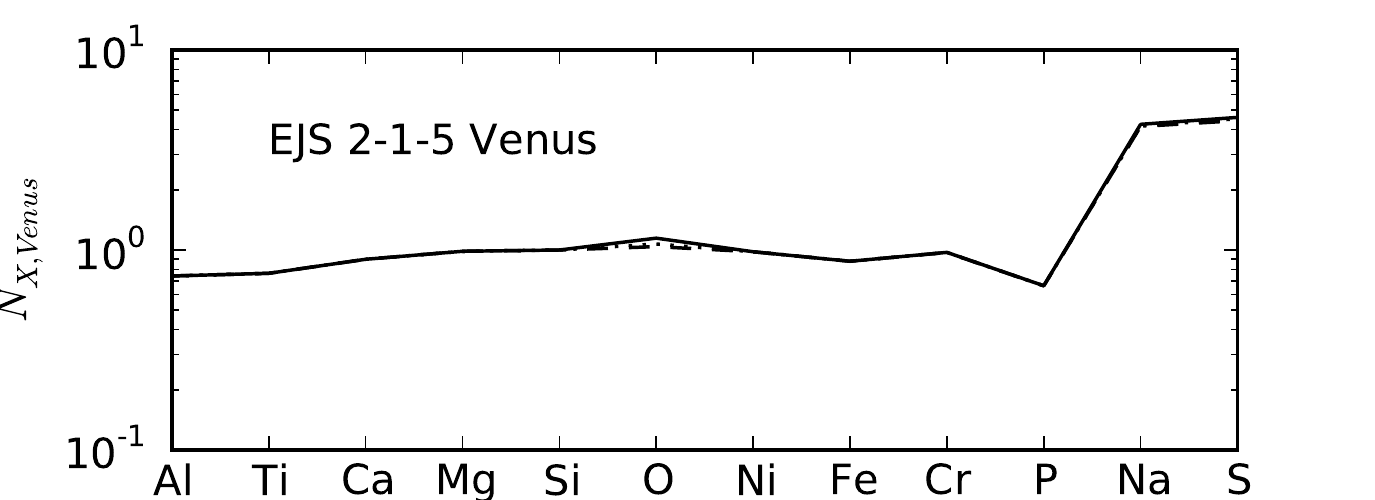}\,\,\,\includegraphics[scale=0.5]{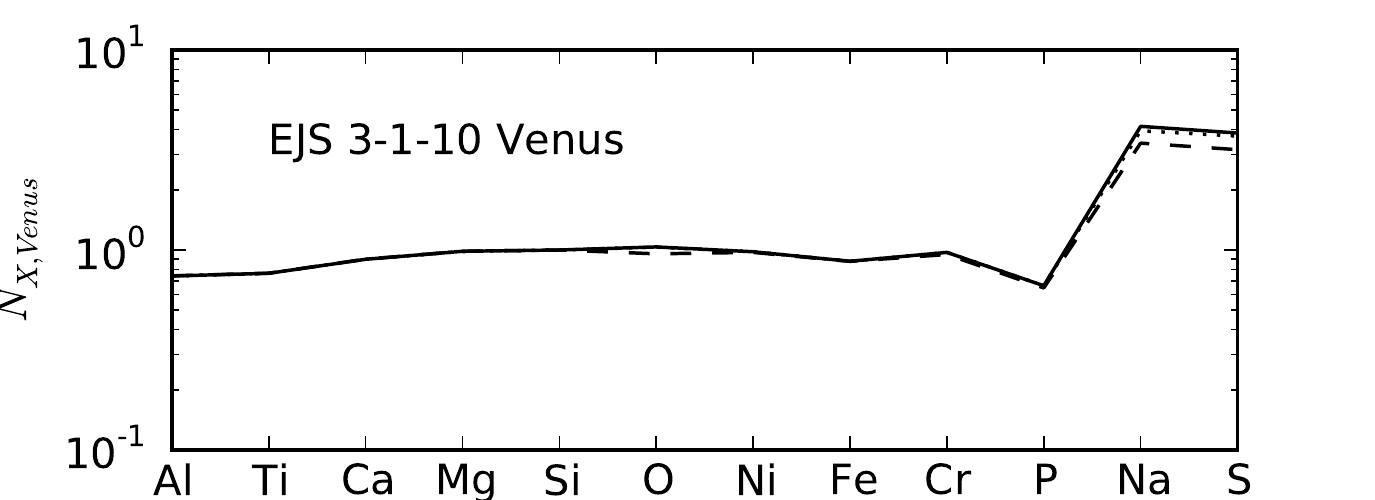}\newline
\includegraphics[scale=0.5]{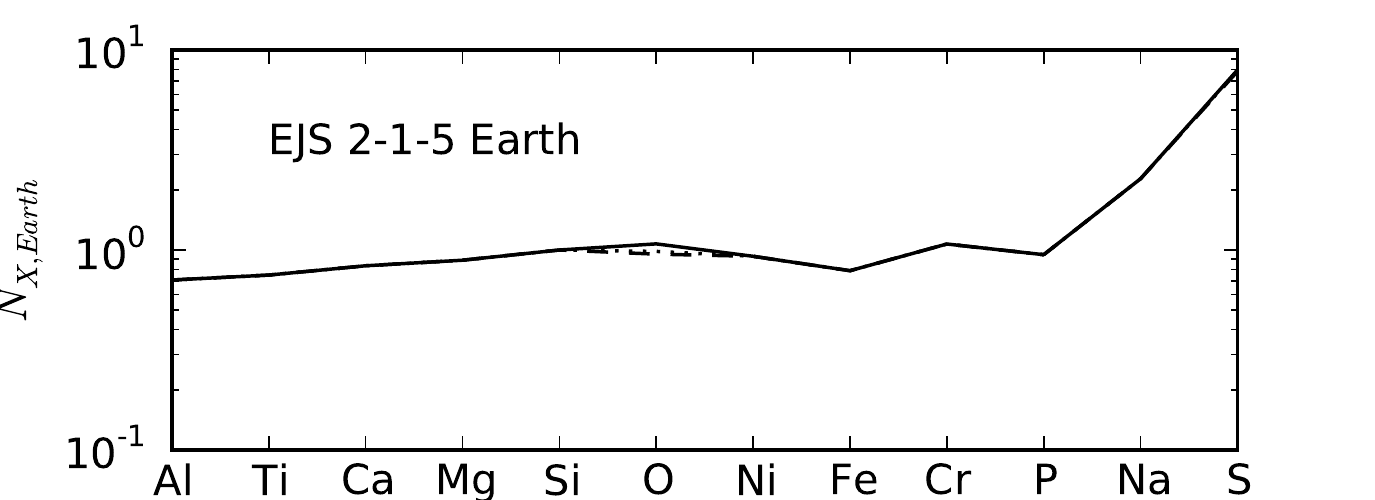}\,\,\,\includegraphics[scale=0.5]{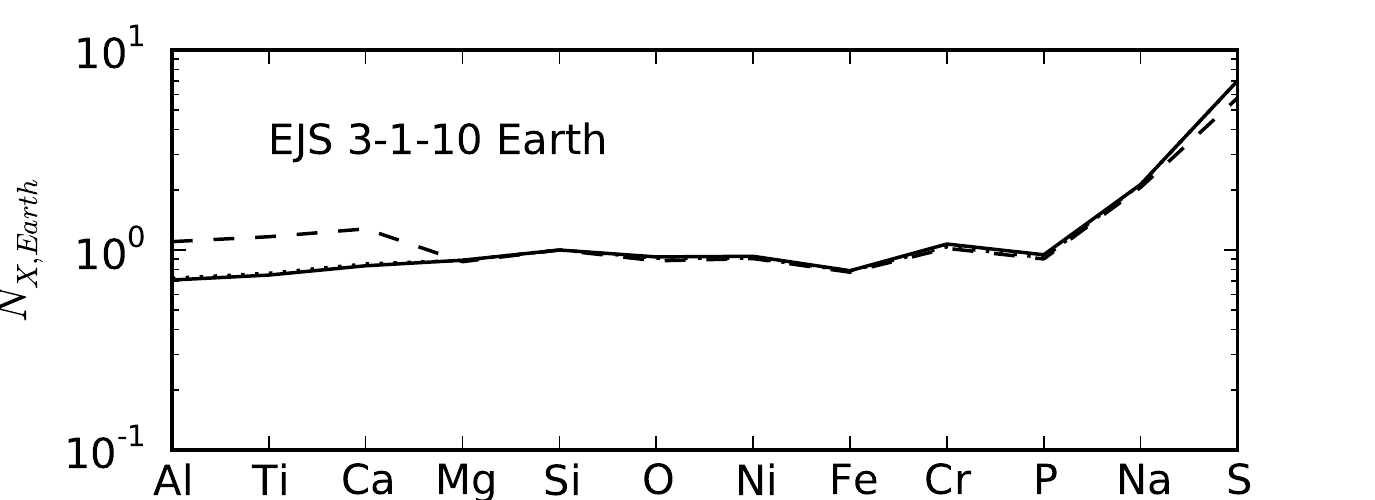}\newline
\includegraphics[scale=0.5]{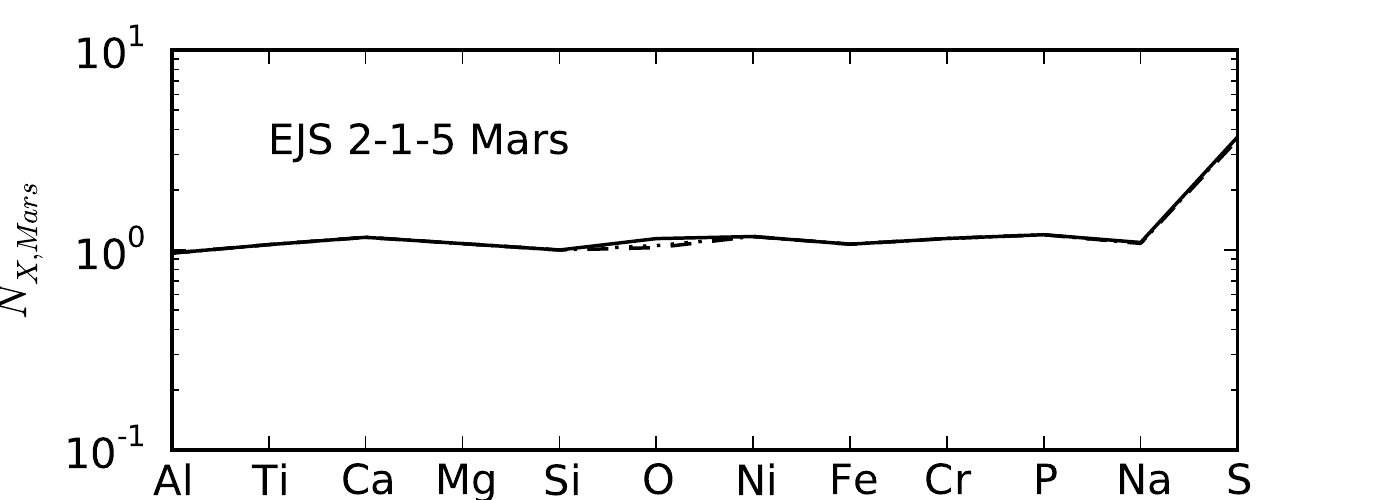}\,\,\,\includegraphics[scale=0.5]{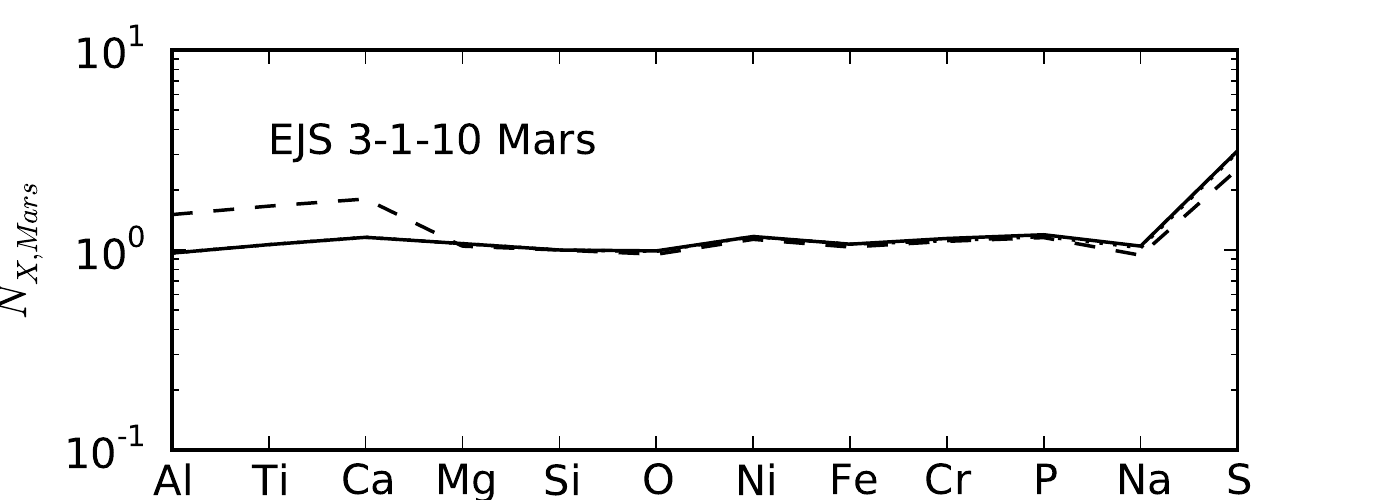}\newline
\caption{The normalized abundance of elements for Mercury, Venus, Earth and Mars analogues, from bottom to top. The elements are listed in order of increasing volatility from left to right. Two-dimensional model (solid line), Chambers model (dashed line) and self similar solution model (dotted line) disks are taken into account. Left column: EJS 2-1-5, right column: EJS 3-1-10.}
\label{fig:NX}%
\end{figure}

In figure \ref{fig:NX}, the normalized abundance $N_{\rm X,tp}$ is shown for each element and all Solar System planets. Both systems fail in reproducing in detail the bulk chemical composition of the Solar System. The most refractory elements (Al, Ti and Ca) are underestimated in Mercury, Venus and Earth analogs. There are two reasons for this: Gas drag and type-I migration leads to the loss of refractory rich planetesimals that form in the inner part of the disk. Hence, they can not be embedded in the planets. On the other hand, if the temperature of the disk is high enough so that refractory solids dominate the solid abundance not only in the very center of the disk, more planetesimals are dominated by refractory material. The Chambers model in the m10 case predicts the highest overall temperature and the normalized abundances of refractory material for the Earth and Mars analogs are significantly increased. The most volatile element abundances (Na and S) are overestimated. This discrepancy is higher if the disk is too cold. In this case volatile rich planetesimals dominate the disk and the normalized bulk abundance becomes too high in the case of all four planets. 
Except in the case of S, the composition of Mars is reproduced by most of the disk models. On the other hand, Mercury is not very well reproduced in any of the models. 

\begin{figure}

\centering
\includegraphics[scale=0.7]{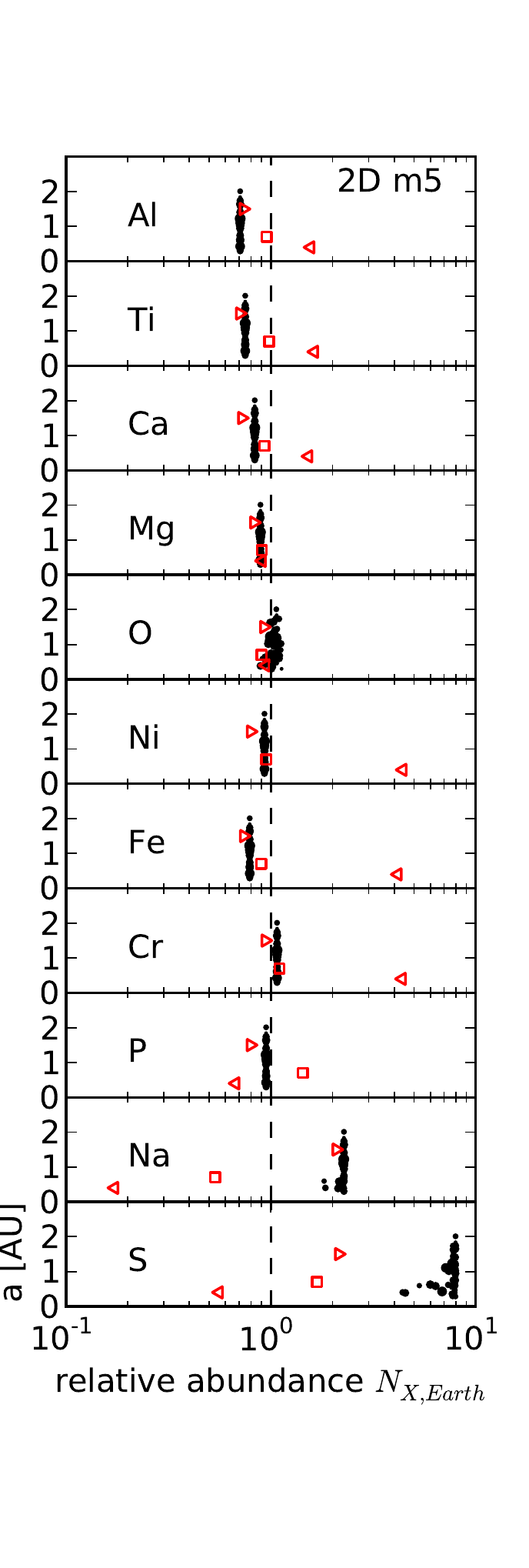}\,\,\,\includegraphics[scale=0.7]{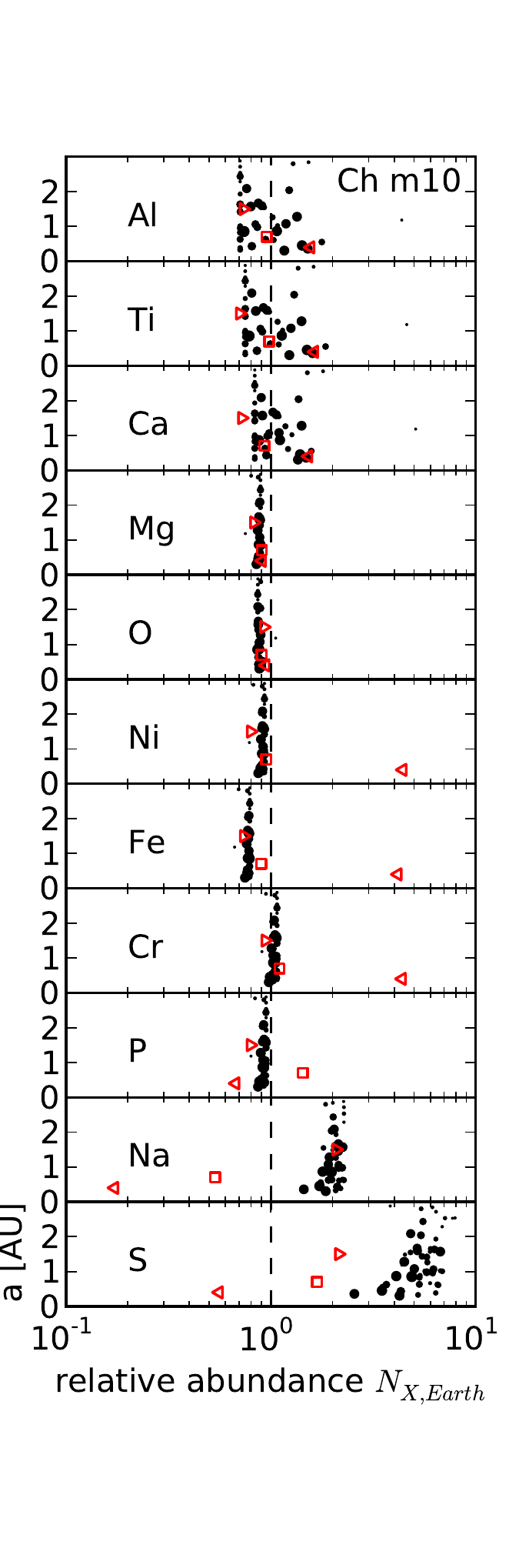}
\caption{The element abundance of all planets. The ordering of the elements increase in volatility from top to bottom. The horizontal axis gives the abundance of element X of a planet relative to the abundance of Si, see (\ref{eq:Nx}) normalized by Earth values. The vertical axis gives the semi-major axis of the planet. The diameter of the black dots represents the mass of the simulated planets. The red markers indicate location of Mercury (left triangle), Venus (square) and Mars (right triangle), the dashed line highlights the Earth abundance. The left column represents m5 with a two-dimensional disk model (2D), this is the model that predicts the lowest disk temperature. The right column represents m10 with a Chambers disk model (Ch), this is the model that predicts the highest disk temperature.}
\label{fig:abundanceALL}%
\end{figure}

Note that the differences in bulk composition of the planets we are considering is small. In both simulations, the abundances differ by up to $50\,\%$ in the case of refractory elements if the Chambers-model is used. All the other element abundances nearly coincide. The fact that the variation in bulk chemical composition in the simulated planets is in general smaller than in the Solar System rocky planets is also demonstrated in figure \ref{fig:abundanceALL}.

Terrestrial planet formation and especially the chaotic growth phase when many giant impacts take place is a stochastic process. This might result in the formation of unique planets that reproduce the bulk chemical composition of certain Solar System planets very well, but to reproduce the bulk composition of all Solar System planets in a single simulation might be a very rare situation. Hence, we study the distribution of the element abundance of all planets at once to get an overview as to how extreme the composition of planets in those simulations can be. As source regions of different planets can differ significantly, we expect variations in the element abundances depending on the underlying disk model. In figure \ref{fig:abundanceALL}, the distribution of the elemental bulk chemical compositions of planets is shown for each element in the case of the most extreme disk models normalized to the Earth values. One model is the two-dimensional model in the case m5. This is the model that gives the lowest temperature across the disk and a lot of species condense across the planetesimal annulus. Thus, it results the most inhomogeneous element abundance profile. The other model is the Chambers model in the m10 case where the resulting element abundance profiles are the most featureless since this disk has the highest temperature. Although these models are applied one two different samples of planets, comparing their element abundances reveals the most extreme variations that are possible in our approach.

In the two-dimensional model, the relative abundances of the simulated planets are mostly located in a narrow range. In the case of refractory elements, the abundances are a little lower than the Earth value and they agree with Mars. There is a variation in O, especially for small semi-major axes. In the case of the volatile elements Na and S, the bulk chemical composition is very high relative to the Earth values, up to a factor of 9 in the case of sulfur. Some planets with small semi-major axis tend to have a smaller abundance of these volatile elements.

In the Chambers model, there is a variation in the relative abundances,  up to a factor of 5 in the case of sulfur. Concerning refractory elements, some planets still group along a narrow range at the same position as in the two-dimensional model. They form a lower bound of the distribution. Some planets have much higher element bulk abundances and agree with the relative abundances of Mercury. Beside some outliers, these planets have small semi-major axes. The volatile element abundances are reproduced in a similar manner than before, although the variation towards lower relative abundances is a little larger. 

The bulk chemical compositions are much more sensitive to the different source regions than in the cooler two-dimensional model. This can be explained using the example of sulfur: in the two-dimensional model FeS condenses around 1 AU, hence most planets incorporate planetesimals with a similar weight percentage of sulfur. The few outliers result from the difference in the source regions. In the Chambers model, condensed sulfur is only present outside 2.5 AU. Hence, depending on the source region, a larger variation in the bulk composition is possible.  

Another example is given by the refractory element Al. In the hot disk, Al is the dominant element inside the condensation radius of more volatile solids at around 1.3 AU (see figure \ref{fig:chemradius}). Hence, planetesimals that are locate initially inside this radius have a wt.$\%$ that is higher than any planetesimals that form in the colder disk. This results in the higher variation.

In both models, the differences between the simulated planets and the Solar System planets are significant, especially in the volatile regime and in the case of Mercury. 

Geochemical ratios allow us to quantify the depletion in volatile elements (e.g. Na/Si) with respect to the enrichment in refractory elements (e.g. Al/Mg). The ratios in the Solar photosphere (table \ref{tab:element}) are Na/Si = 0.045 and Al/Si = 0.072. For the Earth, according to table \ref{tab:abundances}, Na/Si = 0.017 and Al/Si = 0.098. This reflect the volatile depletion and refractory enrichment in the Earth relative to Solar System bulk compositon \citep{Palme03,Rubie11}. Figure \ref{fig:abundanceALL}, which gives the ratio normalized by the reference Earth ratio, shows that the depletion in volatiles is not achieved in any of the simulated planets. Most planets have a Solar photosphere ratio. On the other hand, the suitable enrichment in refractories is provided by some planets in the warm disk model. 
 
\subsection{Water delivery}
In the dynamical simulations by \cite{OBrien06}, the EJS and EEJS simulations result in a water content of Earth-analogs that is very low, since with high giant planet eccentricities the supply of planetesimals from the icy asteroid belt region are suppressed. With the same giant planet orbits, in the simulations created by \cite{Morishima10}, a larger fraction of the planets originates from the outer region of the planetesimal annulus, which should result in a overabundance of water.

Since water is one of the most important molecules in the formation and evolution of life, we focus briefly on the H abundance in the simulated planets to quantify the above statement by \cite{Morishima10}. We keep in mind that the condensation calculations will not be valid at temperatures where hydrated species form. Nevertheless, the abundance of H within a planet in our model will provide a benchmark on the abundance of material from the outer edge of the planetesimal disk.

\begin{figure}
\centering
\includegraphics[scale=0.5]{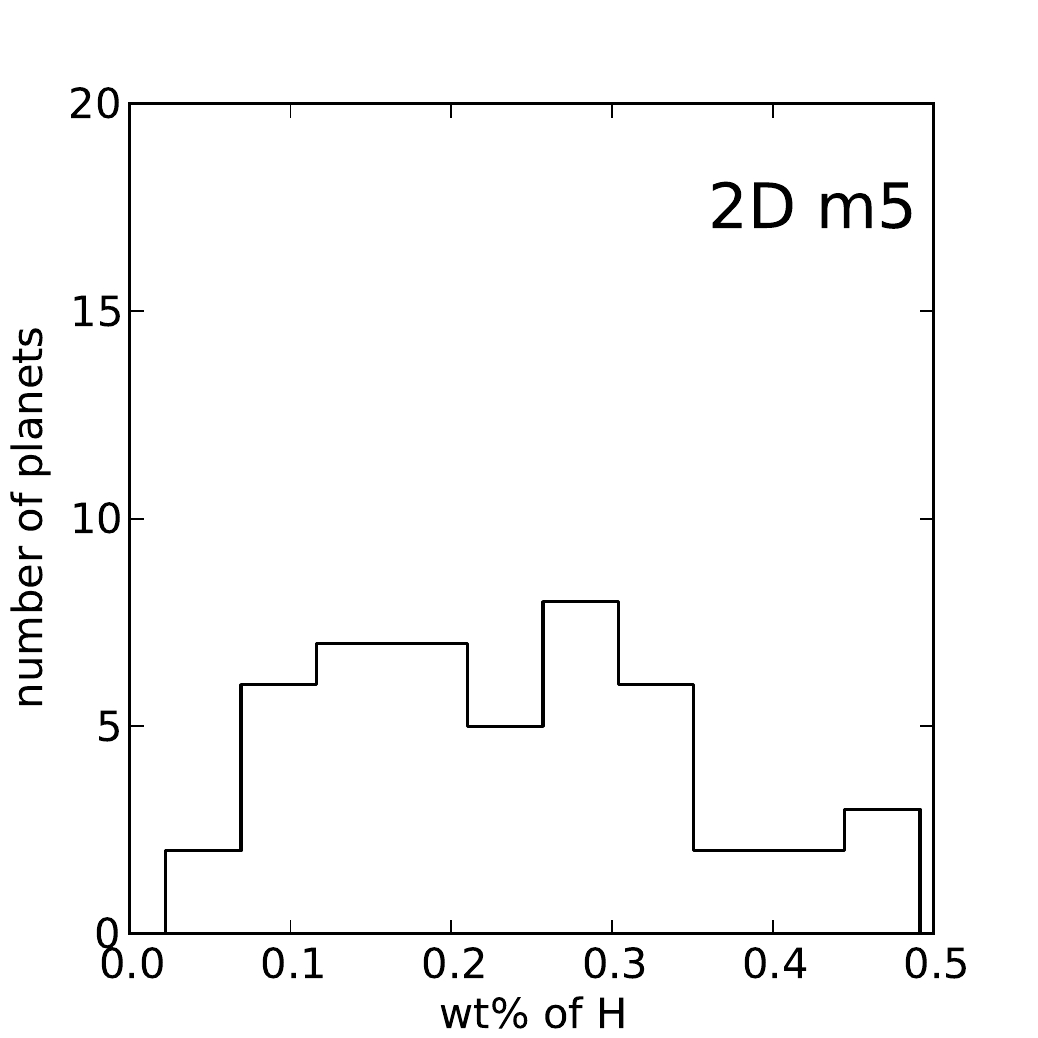}\,\,\,\includegraphics[scale=0.5]{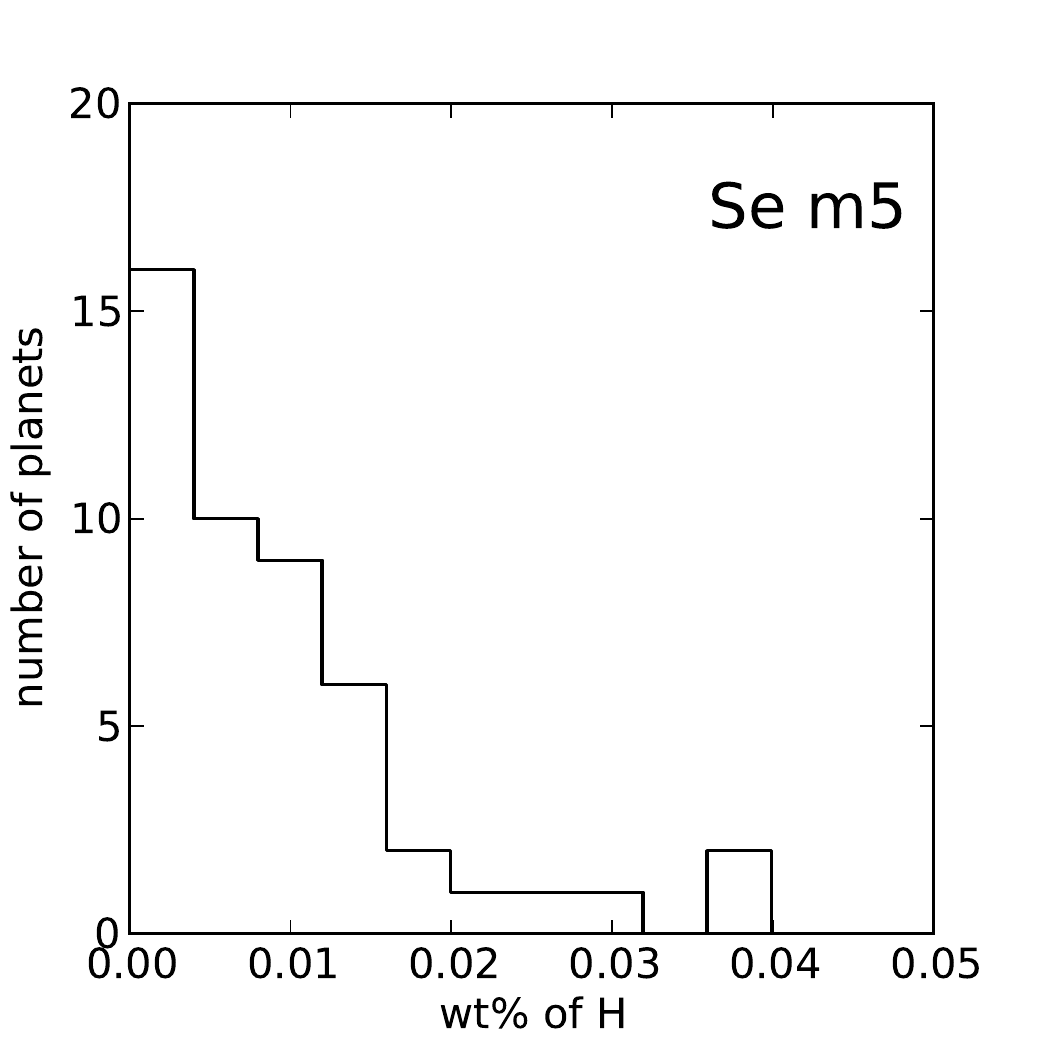}
\caption{The wt.$\%$ of H of all planets for the two-dimensional (2D) and the self similar solution (Se) disk models for m5 disks. All the other models give no condensation of H rich material in the disk. }
\label{fig:Habundance}%
\end{figure}

The H abundance is mainly controlled by the condensation of serpentine. All disks are too hot for water ice to condense. Serpentine condenses out of the solar nebula in the outer region of the planetesimal annulus within 4 AU before $t_{\rm trans}$ depending on the disk model. This edge is located inside 4 AU only in the self-similar solution model and in the two-dimensional model. In figure \ref{fig:Habundance}, the number of planets comprising a wt.$\%$ of H is shown. In general, different disk models lead to a completely different wt.$\%$ of H since its inner edge of condensation is close to the outer edge of the dynamical simulations.

\subsection{Extreme disk chemistry}
In \cite{Bond10a}, the bulk chemical composition of hypothetical terrestrial planets in extrasolar planetary systems is simulated. The element abundance of stars can be significantly different compared to the abundance in our solar nebula and giant planets in other systems can have completely different masses and orbits than Jupiter and Saturn. These circumstances could lead to the formation of rather extreme objects such as carbon-rich planets.

\begin{figure}
\centering
\includegraphics[scale=0.5]{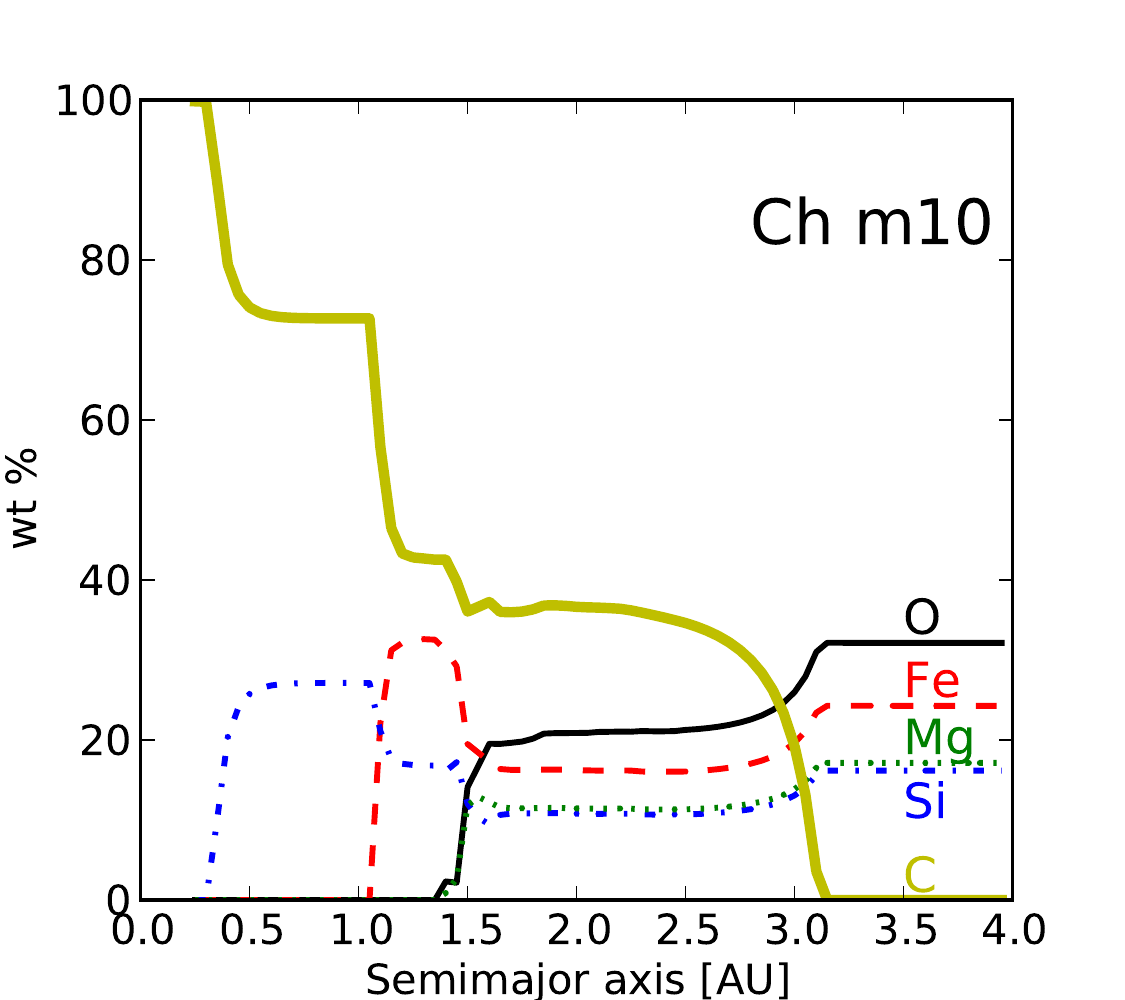}\,\,\,\includegraphics[scale=0.5]{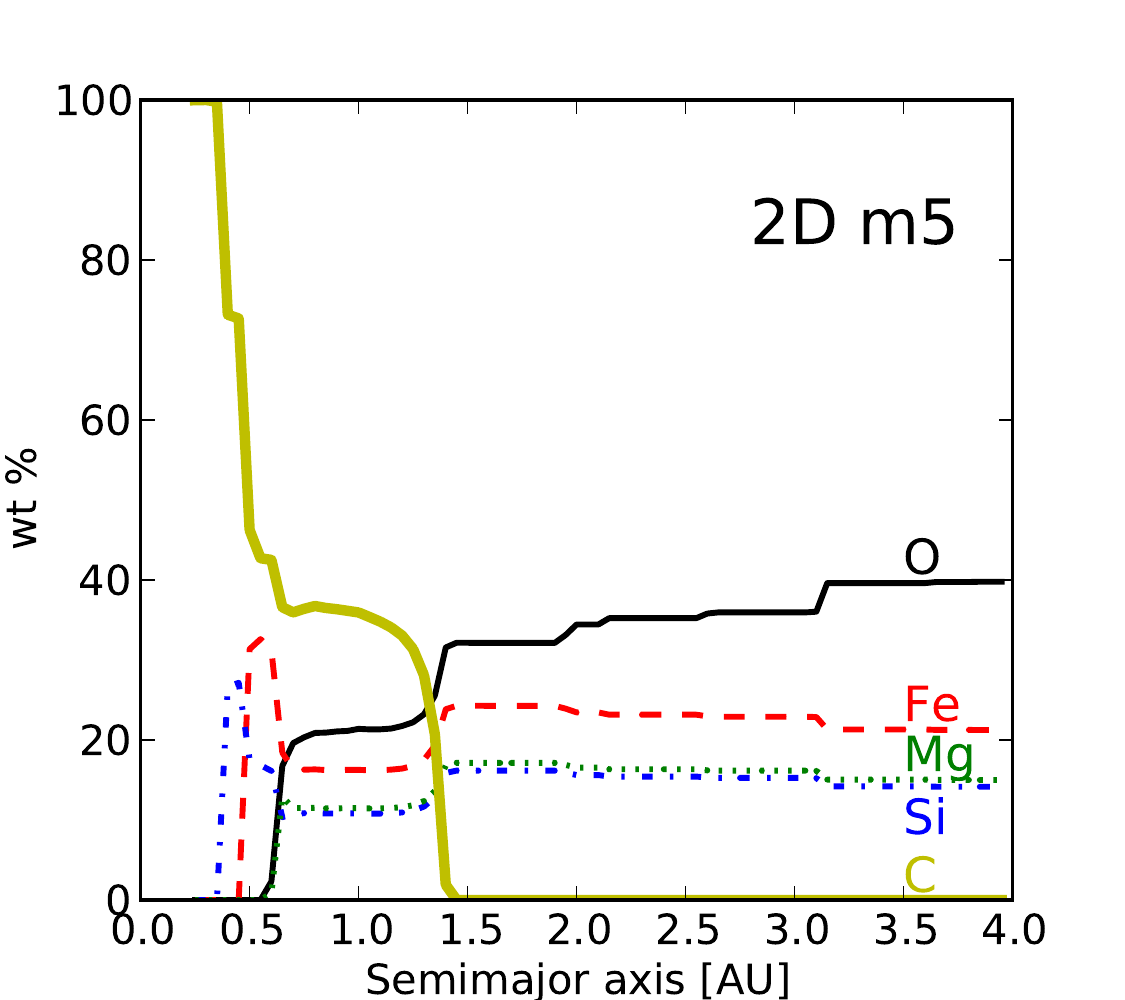}
\caption{These figures show the weight percentage chemical abundance of the most frequent elements in solid species as a function of radius in the case of extreme initial abundances. The models are from left to right: Chambers with m10 and two-dimensional with m5. The thick line shows a very high abundance of C close to the star. At 4 AU, the elements are from top to bottom: O, Fe, Si, Mg, C.  }
\label{fig:abundanceE}%
\end{figure}

\begin{figure}
\centering
\includegraphics[scale=0.7]{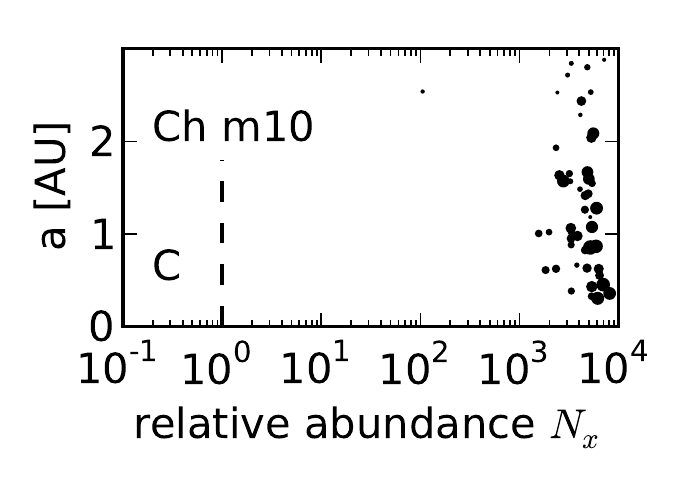}\,\,\,\includegraphics[scale=0.7]{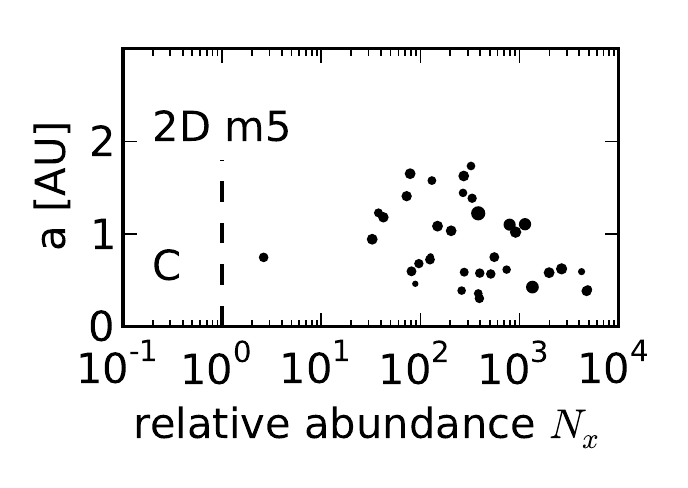}
\caption{The C-abundance of all planets in the case of a extreme solar nebula. Details are explained in the caption of figure \ref{fig:abundanceALL} (no reference values of Mercury, Venus and Mars are shown here). Note the different scaling of the abscissa since the variations are much stronger. The left panel represents m10 with a Chambers disk model (Ch). The right panel represents m5 with a two-dimensional disk model (2D).}
\label{fig:CabundanceE}%
\end{figure}

We explore one extreme case by adopting element abundances of HD 4203 used in \cite{Bond10a}. A recent article by \cite{Fortney12} pointed out that the derived stellar C/O ratios could be overestimated. Nevertheless, we use the same disk models and dynamical simulations as before. If we focus on the most extreme disk models again, figure \ref{fig:abundanceE} reveals that the wt.$\%$ of C becomes extremely high for small radial distances. The main forms of carbon at these distances are C, TiC and SiC. Considering that most of the planetesimals are initially located inside $\sim 1$ AU, planetesimals that will end up in a planet do not have concentrations higher than 40 wt.$\%$. From this, we can expect that a wt.$\%$ of 40 might be the highest bulk abundance that can be achived in the planets. The C-abundance of the planets normalized to the Earth value is shown in figure \ref{fig:CabundanceE}. Note that around one quarter of all planets do not contain C and are not plotted. This also holds in the case of the two-dimensional disk model. In both disk models, there is a wide spread in the C-abundance which can be up to a factor $10^4$ times higher than the Earth C-abundance. Hence, the wt.$\%$ of C in the planets range from a few to 40 percent. This verifies results by \cite{Bond10a}.

\section{Caveats, Implications and Future Work}
\label{sec:Implications}
In this section, we focus on the major caveats and implications that result from our approach concerning the three main goals we introduced at the beginning: First, the estimation and interpretation of abundance profiles in the circumstellar disk. Second, the interpretation of the source regions of the simulated planets in the dynamical simulations. Finally, the combination of abundance profiles and dynamical simulations, the dependence of the resulting bulk chemical compositions on the initial assumptions and an application to the Solar System.

\subsection{Caveats}
\label{sec:caveats}

\paragraph{Disk abundance profiles}

Strong assumptions and simplifications are required when estimating chemical abundance profiles. There are a lot of uncertainties in describing the disk physics correctly, for example the source of viscosity in the disk and the opacity. Solid migration is not taken into account in the disk models, although it is expected to occur in the actual circumstellar disks \citep{Garaud07}. Here, although we used very simple models, they give rise to a broad range of temperature, pressure and surface-density profiles. The self-similar solution model we constructed is not fully self-consistent in its derivation. Since we did not use it in the further study, this caveat is not important. 

All calculations performed with the HSC Chemistry code are made under the assumption of chemical equilibrium.  Each solid species is treated as a separate phase, and each is considered a pure substance. Therefore, only the most stable species appear as condensates. Actually, the formation of solids can be controlled by disequilibrium condensation, at least partially.  In such a case, once a solid has formed, it does not interact anymore with the system. This may occur if condensation takes place fast enough and could have a significant effect on the estimated bulk composition \citep{Bond10}. Other disequilibrium effects are evaporation or photo-dissociation of gaseous molecules \citep{Ebel06}. At low temperatures, the assumption of chemical equilibrium is not valid anymore due to lack of diffusion in the condensed solids. Hence, we can not make reliable predictions for the volatile content in solids and the condensation of hydrated silicates like serpentine, the only solid containing H in our calculations, might be not justified. However, condensation of some hydrated material can not be excluded at the very outer edge of the planetesimal belt. Beside this, the radial abundance of biologically important elements like H, C and N can not be estimated correctly with our approach. C and N rich species do not condense in any of the models. They would condense out outside of 4 AU, were the disk is cold enough. In addition, the formation of water out of H and O depends on its physical environment and the presence of H on a planet need not indicate the presence of $\rm H_2O$. 

The transition from the disk models to the dynamical simulations contains uncertainties. Our transition approach is based on matching the initial surface density of solids at 1 AU by adapting the time of the disk model evolution. 
The slope of the dynamical simulation surface density is exactly reproduced in the self-similar model. The other models provide a flatter profile and the surface density profile deviates from the initial conditions in \cite{Morishima10}.  We did not study the dependence of the abundance profiles on the transition criterion systematically. If the surface density at 1 AU is changed, as it is the case for the two different solid disk masses in the dynamical simulations, the temperature and pressure profiles and thus the element abundance in the disk change significantly. A similar effect occurs when the position of the normalization is changed. If the transition time is left as a free parameter (as in \cite{Bond10}), the abundance profile and thus the bulk compositions will change in a similar manner as if the planetesimal disk mass is changed. If the transition takes place later, the temperature is lower which corresponds to a smaller disk mass. 

Another caveat is the gas-to-dust ratio we adopted to estimate the surface density of solids in our disk models. The \cite{Morishima10} simulations use a gas-to-dust ratio which is up to 3 times higher. This high ratio would result in a very high surface density at $t_{\rm trans}$. The temperature and pressure in the disks would be very high, which moves $r_{\rm in}$ to a larger radial distance. Thus, no solids would be present in large parts of the circumstellar disk at $t_{\rm trans}$. This would be inconsistent with the initial condition of the dynamical simulations. Therefore, we claimed that a gas-to-dust ratio similar to the ISM is more reasonable. Although this results in an inconsistency with the initial conditions of the dynamical simulations, a smaller gas density in the dynamical simulations would only affect the gas-drag and type I migration. The strength of the effect of this type of migration is still a matter of debate \citep{Schlaufman09, Paardekooper08}.

Both caveats concerning the transition mentioned above could be eliminated by adding a more sophisticated disk model to the dynamical simulations which could directly provide temperature and pressure profiles and such a transition would be eliminated.

\paragraph{Source regions of planets}

Observations reveal that there is a very large variety in the configuration of extrasolar systems. Due to limitations in computational resources, the simulations by \cite{Morishima10} were constructed in order to reproduce the constraints given by the Solar System and their outcomes do not represent general planetary systems. The presence of Jupiter and Saturn truncates the initial planetesimal disk at 4 AU and no icy bodies beyond this limit are taken into account. \cite{Schlaufman09} have shown that the type-I migration rate obtained from linear theory decreases by an order of magnitude, if for example the ice line is taken into account. These recent results are not included in the dynamical simulations. In addition, the analytic expression for type-I migration used by \cite{Morishima10} does not account for the sharp surface density transition which may occur at the inner edge of the gas disk, where the disk is truncated by e.g. magnetospheric accretion. Since these strong density gradients can act as protoplanet traps \citep{Masset06,Hasegawa11}, they can prevent early forming protoplanets from falling into the star. Moreover, at the pressure maximum at the inner gas disk edge, gas drag will accumulate planetesimals and the edge can also act as a planetesimal trap. These effects might have significant consequences on the source regions and the final location of the planets. 

Since only one simulation per set of initial conditions was performed by \cite{Morishima10}, we are not able to do a broad study of the dependence of the source regions on the initial conditions of the dynamical simulations. The late stage of formation is a stochastic process and the source regions of two planets in one group might naturally differ significantly between each other. A much larger set of simulations is necessary to explore stochasticity and parameter space.

\paragraph{Combining chemistry and dynamics}

Our transition approach is based on matching the initial surface density of solids at 1 AU by adapting the time of the disk model evolution. This should not imply that all the planetesimals in the dynamical simulations formed instantaneously at their initial positions. Yet, all planetesimals are allocated the composition of solids in the disk only at the time of transition. Actually, the formation and growth of planetesimals out of solids might significantly depend on the local density, temperature and pressure and does occur inhomogeneously along radial direction in the disk. A more sophisticated chemical model could provide condensation rates of solids, so that one could keep track of the solid fraction at every radius and integrate them over time to improve our snapshot-like approach.  

The inner edge $r_{\rm in}$ of the disk of solids that condenses at $t_{\rm trans}$ lies inside the inner edge of the initial planetesimal annulus in the m10 simulations for all disk models. Thus, no solids can be allocated to these planetesimals. However, in all of the dynamical simulations, all the planetesimals initially located inside $r_{\rm in}$ are lost to the star due to gas drag and type I migration and the bulk composition of the final planets can be completely estimated in every case. Nevertheless, two caveats remain: a smaller migration rate \citep{Paardekooper08, Schlaufman09} could prevent some of the innermost planetesimals from migrating into the central star. In addition, the gravity of the planetesimals initially located inside $r_{\rm in}$ gravitational interact with their counterparts on larger radial distance. Hence, they can not simply be ignored. 

According to \cite{Albarede09}, the inner part of the Solar System was very hot as long as gas was present and any solids that condensed were depleted in volatiles relative to the Solar photosphere. Our disk models do not predict such high temperatures at transition time and the expected volatile depletion in the case of e.g. the Earth is not achieved. However, at the inner edge of the planetesimal annulus, highly volatile depleted planetesimals form but most of them are lost into the star. As previously mentioned, modifications in the migration mechanism might result in a higher fraction of volatile depleted planetesimals that end up within planets. Following \cite{Albarede09}, the local planetesimals were dry and the water has to come from asteroids and comets from the Jupiter-Saturn region or comets from the trans-Uranian region \citep{Morbidelli00}. However, water-vapor absorption on silicate grains can be a mechanism to form water-rich planetesimals in the hot regions of the disk \citep{Muralidharan08}. 

The reference values for the bulk composition of the Solar System rocky planets were not reproduced in detail. Since these values are themselves based to some degree on simulations and condensation calculations, they do not necessarily provide real constraints. The bulk composition of Earth \citep{Kargel93} is based on Best Bulk Silicate Earth (BBSE) abundance values, which are estimated using several methods that have their own uncertainties. Knowing the BBSE and the volatility trends for the elements, the core abundances and the bulk chemical composition of the Earth was estimated. Uncertainties in the BBSE values, the volatility trends and the ratio of core mass to total mass result in relative errors from 5-10$\,\%$ for most elements and up to 25$\,\%$ for volatiles. In \cite{Lodders97}, element abundances for Mars are based on meteroite samples and an oxygen isotope model. The uncertainties result in a relative error of 10$\,\%$. In \cite{Morgan80}, since the data is very limited, the compositions of Venus and Mercury are determined fully theoretically. In their method, Solar gas cools under the assumption of equilibrium condensation. The composition of a body is determined by the amount of early condesates which were most likely embodied in the final planet \citep{Morgan78}. The uncertainties arising from this method are not quantified. In summary, the fact that we can reproduce parts of the characteristica of the chemical compositions of the planets does not imply definitly that our model is reproducing the Solar System planets. The reference values should be seen as a guideline. 

There are some further caveats resulting from the dynamical simulations: the limited number of simulations does not provide any planetary system that accurately mimics the inner Solar System. Since terrestrial planet formation is a stochastic process, a larger set of simulations could provide better candidates. The simulations adopt perfect sticking if two bodies merge, and no decrease of volatiles is taken into account. More sophisticated accretion models \citep{Stewart09} propose a collision outcome depending on the collision parameters concerning energy loss and fragmentation. It ranges from a perfect accretion event to complete disruption of impactor or target. The energy release of these events is clearly high enough that volatile elements can be lost. This can be a possible explanation of the overabundance of volatile elements \citep{Bond10}, although there are other additional processes that can result in the depletion in volatiles in the solar nebula \citep{Davis06}. Finally, we note that the rather extreme composition of Mercury, e.g. its massive core, can be explained by a collision that striped off most of Mercury's mantle and crust. No such event can take place in the \cite{Morishima10} simulations.

\subsection{Implications}

\paragraph{Disk abundance profiles}

Different initial conditions in the dynamical simulations result in different abundance profiles in the disk.  Since the temperature and pressure in a disk is directly related to the surface density, the choice of the initial disk mass and the profile of the planetesimals in the dynamical simulations controls the distributions of condensed solids in the disk. Relatively cold disks shelter a smaller variety of solids and especially smaller abundances of volatiles like H. A massive disk of planetesimals such as the m10 simulations implies a hot disk where only a few solids will condense close to the center of the disk and no H is embodied in solids inside 4 AU. A less massive disk like the m5 disk can result in the condensation of the entire range of considered elements across the disk. The initial disk mass has an equally large or even larger effect on the resulting abundance profile than the choice of the disk model.

\paragraph{Source region of planets}

Comparing the source regions of the planets in the different groups reveals that most of them do not coincide. The distributions are sometimes more peaked or more flat, but the width of the source region is always very similar. Due to migration and secular resonances, the final planets are located in the inner part of their source region, some of them are even found outside of their source region. The innermost planets tend to have source regions that are flatter. A trend is visible in the planet mass dependence of source regions: Massive planets have a flat and wide source region while small planets have a steep source region which becomes steeper at larger radial distance. This implies that massive planets consist of elements that condensed from regions that spanned a larger range of physical conditions than their less massive counterparts.

\paragraph{Combining chemistry and dynamics}

The bulk chemical composition of planets produced in the dynamical simulations do not reproduce the bulk chemical composition of the rocky Solar System planets in detail. We studied two realizations of a four-planet system: In both cases, the effect of different disk models or the initial composition was not dramatic and the two dynamical simulations also gave similar results. Thus, it may be difficult to reproduce the diversity in the reference bulk chemical compositions of our Solar System in a single simulation.  

In \cite{Bond10}, the transition time is a free parameter and a different parameter space is covered than in our study. We believe that our transition criterion is more self-consistent, but both methods result in a broad range of planetary compositions.  Similar to \cite{Bond10}, we overestimated the abundance of volatile elements, implying that volatile loss can not be ignored. Keeping in mind the differences in the dynamical simulations by \cite{Morishima10} and \cite{OBrien06}, we found that it is the choice of the disk model (a cold disk) that controls the amount of wet planetesimals that are accreted (rather than the orbits of the Jovian planets).

A comparison of the bulk chemical composition of all simulated planets reveals the effect of the choice of the disk model. If the initial disk mass of the dynamical simulations is small (m5) and the coldest disk model is chosen (two dimensional model), most of the simulated planets share the same bulk composition. Variations are only visible in O, Na, and S for planets with small semi-major axis. In this scenario, Solar System values are almost never achieved. This means that the dynamics are not significantly affecting the abundances of this model. On the other hand, a massive disk (m10) with a hot disk model (Chambers model) provides large variations in the most refractory and volatile elements, since dynamics are important in this scenario. Again, small semi-major axis planets tend to vary more than others.

A similar trend is visible in the case of planet mass, especially for S, since more massive planets vary more in bulk composition than smaller planets. A difference was also observed for the source regions of large mass and small mass planets. Massive planets tend to be comprised of a more uniform distribution of planetesimals that formed all across the disk. Therefore they tend also to vary more in bulk composition than small mass planets, which are comprised of planetesimals from the outer disk region. A larger set of dynamical simulations is needed to quantify this difference. Note that such correlations between mass and bulk composition are not visible in the extreme abundance plot for C (figure \ref{fig:CabundanceE}). The C abundance dominates over a larger region than for example Al in the case of a Solar nebula composition and thus, the different source regions of small and large mass planets do not play an important role. 

Both the self-similar solution model and the two-dimensional disk model in case of the less massive disk, give rise to the condensation of H rich material inside the planetesimal region. Water is an important component in the formation and evolution of life as we know it. If predictions about water on habitable planets in such simulations are made, the choice of the disk model and the initial disk mass have a significant effect on the amount of H that is delivered to the planets. However, condensation calculations fail in this temperature regime and perhaps the planetesimals form dry and water has to be delivered from outside the planetesimal belt.

\subsection{Future work}

Given all of the uncertainties and assumptions, the main caveat in this work is the limited number of dynamical simulations that were available. A new code currently under development at the University of Zurich that will be able to perform terrestrial planet formation simulations using GPUs and a much larger number of simulations is planned to explore stochasticity and a broader parameter space. With a more sophisticated disk model in the dynamical simulations, estimating the bulk composition of planetesimals should be possible more accurately and self-consistently. A large set of dynamical simulations will also provide a good starting point for statistical studies on the source region of planets. Future studies should also include a more detailed treatment of the condensation sequence which would provide a better understanding of the evolution of solid abundances in the disk.

\section{Conclusions}

\label{sec:Conclusions}

The combination of chemical disk models and dynamical simulations opens a new avenue for exploring planet formation models and to interpret the history of the terrestrial planets \citep{Bond10}. The output of the coupled disk model, its chemistry and the dynamical simulations leads directly to estimates of the bulk composition of the simulated planets. We explored the sensitivity of the radial element abundance trends in the resulting planets to the disk model assumptions, the initial conditions of the dynamical simulations and initial composition of the solar nebula.
Our main conclusions are:

\begin{itemize}
	\item \cite{Bond10a} concluded that their estimated bulk chemical composition are in excellent agreement with observed planetary values, indicating that the models in the dynamical simulations of \cite{OBrien06} work properly in reproducing the bulk composition of the inner Solar System. In our study, these compositions are not reproduced in detail, resulting from the different dynamical simulations and the more restrictive but self-consistent transition from the gas disk to the planetesimals.  The largest discrepancies are in the abundances of volatiles and refractories and in the bulk composition of a Mercury analogue.
	\item The source regions of planets are unique. Large mass planets form from a broader region of the initial disk whilst lower mass planets tend to comprise of planetesimals which are initially located in the outer region of the disk.
	\item The element abundance profiles change significantly if different disk models or if different initial disk masses of planetesimals are used. Some elements do not condense at all if the disk model predicts a high temperature along the disk. This dependence has a significant effect on the estimated bulk chemical compositions of the planets.
	\item In a cold disk model, the variety in the bulk compositions is small and mixing is not important. In a hot disk model, the dynamics lead to a large variety in the bulk compositions of planets in the case of the most refractory and most volatile elements. The variation tends to be larger for massive planets and for those with small semi-major axis.
\end{itemize}

\section*{Acknowledgments}
We appreciate the comments by the reviewers Franck Hersant and David Rubie. We thank Jade Bond for her inspiring work on terrestrial planet bulk compositions. We thank Pascale Garaud for a useful discussion on the disk models and James Connolly for suggestions on equilibrium calculation software. We thank Ryuji Morishima for performing the dynamical simulations on the zBox supercomputer at the University of Zurich. We thank the Planet-Z community for the fruitful discussions and for providing a platform of collaboration and we thank the University of Zurich for financial support.

\newpage
\section*{Online supplementary material}
Tables \ref{tab7} to \ref{tab12} show the bulk chemical composition of all planets in the simulations for different disk models and disk masses. They are published as online material.
\newpage
\addtolength{\textwidth}{0.3in}

\begin{landscape}
\begin{longtable}{c c r r l  l l l l l l l l l l l l l l l}
\caption{Bulk chemical composition of all planets with a Chambers disk model and $m_T=5\,m_\oplus$. } \\
\hline
orbits & $\tau_{\rm decay}$ & $a [AU]$ & $m_p [m_{\oplus}]$ &Al & H & O & S & C & N & F & Ca & Cr & Fe & He & Mg & Na & Ni & Si & Ti \\
\hline
\endfirsthead
\hline
orbits & $\tau_{\rm decay}$ & $a [AU]$ & $m_p [m_{\oplus}]$ &Al & H & O & S & C & N & F & Ca & Cr & Fe & He & Mg & Na & Ni & Si & Ti \\\hline
\endhead
\hline
\multicolumn{19}{c}{continued on next page } \\
\hline
\endfoot
\hline
\endlastfoot
CJS & 1 & 0.39 & 0.68 &  1.26 &  0.00 & 31.48 &  2.54 &  0.00 &  0.00 &  0.12 &  1.66 &  0.39 & 28.64 &  0.00 & 15.01 &  0.41 &  1.79 & 16.62 &  0.07 \\
CJS & 1 & 0.59 & 0.23 &  1.49 &  0.00 & 31.77 &  2.87 &  0.00 &  0.00 &  0.10 &  1.96 &  0.33 & 28.09 &  0.00 & 14.78 &  0.46 &  1.58 & 16.49 &  0.09 \\
CJS & 1 & 1.02 & 0.93 &  1.10 &  0.00 & 31.02 &  6.06 &  0.00 &  0.00 &  0.12 &  1.45 &  0.39 & 27.36 &  0.00 & 14.32 &  0.57 &  1.73 & 15.81 &  0.07 \\
CJS & 1 & 1.63 & 0.77 &  1.09 &  0.00 & 30.75 &  6.99 &  0.00 &  0.00 &  0.12 &  1.43 &  0.39 & 27.07 &  0.00 & 14.16 &  0.58 &  1.72 & 15.64 &  0.06 \\
& & & & & & & & & & & & & & & & & & & \\
CJS & 2 & 0.62 & 0.39 &  1.10 &  0.00 & 30.87 &  6.14 &  0.00 &  0.00 &  0.12 &  1.45 &  0.40 & 27.40 &  0.00 & 14.34 &  0.56 &  1.74 & 15.83 &  0.07 \\
CJS & 2 & 0.94 & 0.46 &  1.07 &  0.00 & 31.50 &  7.15 &  0.00 &  0.00 &  0.12 &  1.41 &  0.38 & 26.67 &  0.00 & 13.95 &  0.58 &  1.69 & 15.41 &  0.06 \\
CJS & 2 & 1.41 & 0.69 &  1.07 &  0.00 & 31.53 &  7.17 &  0.00 &  0.00 &  0.12 &  1.41 &  0.38 & 26.65 &  0.00 & 13.94 &  0.58 &  1.69 & 15.40 &  0.06 \\
\end{longtable}
\begin{longtable}{c c r r l  l l l l l l l l l l l l l l l}
\caption{Bulk chemical composition of all planets with a two-dimensional disk model and $m_T=5\,m_\oplus$. } \\
\hline
orbits & $\tau_{\rm decay}$ & $a [AU]$ & $m_p [m_{\oplus}]$ &Al & H & O & S & C & N & F & Ca & Cr & Fe & He & Mg & Na & Ni & Si & Ti \\
\hline
\endfirsthead
\hline
orbits & $\tau_{\rm decay}$ & $a [AU]$ & $m_p [m_{\oplus}]$ &Al & H & O & S & C & N & F & Ca & Cr & Fe & He & Mg & Na & Ni & Si & Ti \\\hline
\endhead
\hline
\multicolumn{19}{c}{continued on next page } \\
\hline
\endfoot
\hline
\endlastfoot
CJS & 1 & 0.39 & 0.68 &  1.12 &  0.02 & 31.20 &  4.47 &  0.00 &  0.00 &  0.13 &  1.48 &  0.40 & 27.97 &  0.00 & 14.64 &  0.57 &  1.77 & 16.17 &  0.07 \\
CJS & 1 & 0.59 & 0.23 &  1.11 &  0.07 & 31.43 &  5.16 &  0.00 &  0.00 &  0.12 &  1.46 &  0.40 & 27.59 &  0.00 & 14.44 &  0.48 &  1.75 & 15.94 &  0.07 \\
CJS & 1 & 1.02 & 0.93 &  1.06 &  0.11 & 32.20 &  6.89 &  0.00 &  0.00 &  0.12 &  1.40 &  0.38 & 26.43 &  0.00 & 13.83 &  0.57 &  1.67 & 15.27 &  0.06 \\
CJS & 1 & 1.63 & 0.77 &  1.05 &  0.10 & 32.56 &  7.19 &  0.00 &  0.00 &  0.12 &  1.38 &  0.38 & 26.15 &  0.00 & 13.68 &  0.56 &  1.66 & 15.11 &  0.06 \\
& & & & & & & & & & & & & & & & & & & \\
CJS & 2 & 0.62 & 0.39 &  1.06 &  0.06 & 32.11 &  6.85 &  0.00 &  0.00 &  0.12 &  1.40 &  0.38 & 26.51 &  0.00 & 13.87 &  0.57 &  1.68 & 15.32 &  0.06 \\
CJS & 2 & 0.94 & 0.46 &  1.03 &  0.19 & 33.33 &  7.23 &  0.00 &  0.00 &  0.12 &  1.36 &  0.37 & 25.76 &  0.00 & 13.48 &  0.56 &  1.63 & 14.89 &  0.06 \\
CJS & 2 & 1.41 & 0.69 &  1.03 &  0.26 & 33.68 &  7.16 &  0.00 &  0.00 &  0.12 &  1.35 &  0.37 & 25.61 &  0.00 & 13.40 &  0.55 &  1.62 & 14.80 &  0.06 \\
\end{longtable}
\begin{longtable}{c c r r l  l l l l l l l l l l l l l l l}
\caption{Bulk chemical composition of all planets with a self-similar solution disk model and $m_T=5\,m_\oplus$. } \\
\hline
orbits & $\tau_{\rm decay}$ & $a [AU]$ & $m_p [m_{\oplus}]$ &Al & H & O & S & C & N & F & Ca & Cr & Fe & He & Mg & Na & Ni & Si & Ti \\
\hline
\endfirsthead
\hline
orbits & $\tau_{\rm decay}$ & $a [AU]$ & $m_p [m_{\oplus}]$ &Al & H & O & S & C & N & F & Ca & Cr & Fe & He & Mg & Na & Ni & Si & Ti \\\hline
\endhead
\hline
\multicolumn{19}{c}{continued on next page } \\
\hline
\endfoot
\hline
\endlastfoot
CJS & 1 & 0.39 & 0.68 &  1.25 &  0.00 & 31.38 &  3.17 &  0.00 &  0.00 &  0.12 &  1.65 &  0.39 & 28.40 &  0.00 & 14.87 &  0.44 &  1.78 & 16.48 &  0.07 \\
CJS & 1 & 0.59 & 0.23 &  1.48 &  0.00 & 31.74 &  3.48 &  0.00 &  0.00 &  0.10 &  1.95 &  0.33 & 27.85 &  0.00 & 14.61 &  0.47 &  1.58 & 16.33 &  0.09 \\
CJS & 1 & 1.02 & 0.93 &  1.09 &  0.00 & 31.40 &  6.39 &  0.00 &  0.00 &  0.12 &  1.43 &  0.39 & 27.05 &  0.00 & 14.15 &  0.57 &  1.71 & 15.63 &  0.06 \\
CJS & 1 & 1.63 & 0.77 &  1.07 &  0.00 & 31.34 &  7.11 &  0.00 &  0.00 &  0.12 &  1.41 &  0.39 & 26.76 &  0.00 & 14.00 &  0.57 &  1.70 & 15.46 &  0.06 \\
& & & & & & & & & & & & & & & & & & & \\
CJS & 2 & 0.62 & 0.39 &  1.08 &  0.00 & 31.42 &  6.51 &  0.00 &  0.00 &  0.12 &  1.43 &  0.39 & 27.00 &  0.00 & 14.13 &  0.55 &  1.71 & 15.60 &  0.06 \\
CJS & 2 & 0.94 & 0.46 &  1.06 &  0.00 & 31.89 &  7.31 &  0.00 &  0.00 &  0.12 &  1.40 &  0.38 & 26.43 &  0.00 & 13.83 &  0.57 &  1.67 & 15.27 &  0.06 \\
CJS & 2 & 1.41 & 0.69 &  1.06 &  0.01 & 32.14 &  7.24 &  0.00 &  0.00 &  0.12 &  1.39 &  0.38 & 26.35 &  0.00 & 13.79 &  0.57 &  1.67 & 15.22 &  0.06 \\
\end{longtable}
\begin{longtable}{c c r r l  l l l l l l l l l l l l l l l}
\caption{Bulk chemical composition of all planets with a Chambers disk model and $m_T=10\,m_\oplus$. } \\
\hline
orbits & $\tau_{\rm decay}$ & $a [AU]$ & $m_p [m_{\oplus}]$ &Al & H & O & S & C & N & F & Ca & Cr & Fe & He & Mg & Na & Ni & Si & Ti \\
\hline
\endfirsthead
\hline
orbits & $\tau_{\rm decay}$ & $a [AU]$ & $m_p [m_{\oplus}]$ &Al & H & O & S & C & N & F & Ca & Cr & Fe & He & Mg & Na & Ni & Si & Ti \\\hline
\endhead
\hline
\multicolumn{19}{c}{continued on next page } \\
\hline
\endfoot
\hline
\endlastfoot
CJS & 1 & 0.36 & 1.3 &  2.41 &  0.00 & 31.72 &  2.53 &  0.00 &  0.00 &  0.12 &  2.65 &  0.38 & 27.35 &  0.00 & 14.34 &  0.39 &  1.71 & 16.26 &  0.14 \\
CJS & 1 & 0.62 & 0.65 &  1.51 &  0.00 & 30.97 &  3.65 &  0.00 &  0.00 &  0.12 &  1.68 &  0.40 & 28.19 &  0.00 & 14.76 &  0.51 &  1.77 & 16.35 &  0.09 \\
CJS & 1 & 1.28 & 1.31 &  2.11 &  0.00 & 31.10 &  4.39 &  0.00 &  0.00 &  0.12 &  2.49 &  0.38 & 26.88 &  0.00 & 14.12 &  0.51 &  1.69 & 16.08 &  0.12 \\
CJS & 1 & 2.09 & 1.07 &  1.22 &  0.00 & 30.57 &  4.79 &  0.00 &  0.00 &  0.12 &  1.60 &  0.39 & 27.97 &  0.00 & 14.66 &  0.55 &  1.75 & 16.30 &  0.07 \\
& & & & & & & & & & & & & & & & & & & \\
CJS & 2 & 0.43 & 0.94 &  1.29 &  0.00 & 30.71 &  4.28 &  0.00 &  0.00 &  0.13 &  1.70 &  0.40 & 28.08 &  0.00 & 14.70 &  0.57 &  1.77 & 16.31 &  0.08 \\
CJS & 2 & 1.01 & 0.17 &  1.12 &  0.00 & 30.22 &  5.56 &  0.00 &  0.00 &  0.13 &  1.48 &  0.40 & 27.96 &  0.00 & 14.63 &  0.51 &  1.77 & 16.16 &  0.07 \\
CJS & 2 & 1.57 & 1.29 &  1.24 &  0.00 & 29.97 &  6.52 &  0.00 &  0.00 &  0.12 &  1.59 &  0.40 & 27.47 &  0.00 & 14.37 &  0.59 &  1.74 & 15.91 &  0.07 \\
\end{longtable}
\begin{longtable}{c c r r l  l l l l l l l l l l l l l l l}
\caption{Bulk chemical composition of all planets with a two-dimensional disk model and $m_T=10\,m_\oplus$. } \\
\hline
orbits & $\tau_{\rm decay}$ & $a [AU]$ & $m_p [m_{\oplus}]$ &Al & H & O & S & C & N & F & Ca & Cr & Fe & He & Mg & Na & Ni & Si & Ti \\
\hline
\endfirsthead
\hline
orbits & $\tau_{\rm decay}$ & $a [AU]$ & $m_p [m_{\oplus}]$ &Al & H & O & S & C & N & F & Ca & Cr & Fe & He & Mg & Na & Ni & Si & Ti \\\hline
\endhead
\hline
\multicolumn{19}{c}{continued on next page } \\
\hline
\endfoot
\hline
\endlastfoot
CJS & 1 & 0.36 & 1.3 &  1.14 &  0.00 & 30.98 &  4.63 &  0.00 &  0.00 &  0.12 &  1.50 &  0.39 & 28.02 &  0.00 & 14.66 &  0.55 &  1.75 & 16.20 &  0.07 \\
CJS & 1 & 0.62 & 0.65 &  1.11 &  0.00 & 31.02 &  6.02 &  0.00 &  0.00 &  0.12 &  1.46 &  0.39 & 27.38 &  0.00 & 14.32 &  0.58 &  1.71 & 15.83 &  0.07 \\
CJS & 1 & 1.28 & 1.31 &  1.22 &  0.00 & 31.02 &  6.24 &  0.00 &  0.00 &  0.12 &  1.57 &  0.39 & 27.16 &  0.00 & 14.21 &  0.55 &  1.72 & 15.73 &  0.07 \\
CJS & 1 & 2.09 & 1.07 &  1.09 &  0.00 & 30.91 &  6.65 &  0.00 &  0.00 &  0.12 &  1.43 &  0.39 & 27.15 &  0.00 & 14.21 &  0.57 &  1.72 & 15.69 &  0.06 \\
& & & & & & & & & & & & & & & & & & & \\
CJS & 2 & 0.43 & 0.94 &  1.09 &  0.00 & 30.92 &  6.75 &  0.00 &  0.00 &  0.12 &  1.43 &  0.39 & 27.10 &  0.00 & 14.18 &  0.58 &  1.72 & 15.66 &  0.06 \\
CJS & 2 & 1.01 & 0.17 &  1.08 &  0.00 & 31.48 &  6.55 &  0.00 &  0.00 &  0.12 &  1.42 &  0.39 & 26.94 &  0.00 & 14.10 &  0.58 &  1.71 & 15.57 &  0.06 \\
CJS & 2 & 1.57 & 1.29 &  1.07 &  0.00 & 31.42 &  7.33 &  0.00 &  0.00 &  0.12 &  1.41 &  0.38 & 26.63 &  0.00 & 13.94 &  0.57 &  1.69 & 15.39 &  0.06 \\
\end{longtable}
\begin{longtable}{c c r r l  l l l l l l l l l l l l l l l}
\caption{Bulk chemical composition of all planets with a self-similar solution disk model and $m_T=10\,m_\oplus$. } \\
\hline
orbits & $\tau_{\rm decay}$ & $a [AU]$ & $m_p [m_{\oplus}]$ &Al & H & O & S & C & N & F & Ca & Cr & Fe & He & Mg & Na & Ni & Si & Ti \\
\hline
\endfirsthead
\hline
orbits & $\tau_{\rm decay}$ & $a [AU]$ & $m_p [m_{\oplus}]$ &Al & H & O & S & C & N & F & Ca & Cr & Fe & He & Mg & Na & Ni & Si & Ti \\\hline
\endhead
\hline
\multicolumn{19}{c}{continued on next page } \\
\hline
\endfoot
\hline
\endlastfoot
CJS & 1 & 0.36 & 1.3 &  2.15 &  0.00 & 31.32 &  4.44 &  0.00 &  0.00 &  0.12 &  2.00 &  0.39 & 27.19 &  0.00 & 14.23 &  0.48 &  1.71 & 15.87 &  0.11 \\
CJS & 1 & 0.62 & 0.65 &  1.59 &  0.00 & 31.12 &  5.88 &  0.00 &  0.00 &  0.12 &  1.51 &  0.39 & 27.14 &  0.00 & 14.20 &  0.56 &  1.72 & 15.69 &  0.08 \\
CJS & 1 & 1.28 & 1.31 &  1.10 &  0.00 & 30.73 &  6.18 &  0.00 &  0.00 &  0.12 &  1.45 &  0.38 & 27.22 &  0.00 & 14.25 &  0.53 &  1.71 & 15.74 &  0.07 \\
CJS & 1 & 2.09 & 1.07 &  1.09 &  0.00 & 30.80 &  6.58 &  0.00 &  0.00 &  0.12 &  1.44 &  0.39 & 27.23 &  0.00 & 14.25 &  0.56 &  1.73 & 15.74 &  0.06 \\
& & & & & & & & & & & & & & & & & & & \\
CJS & 2 & 0.43 & 0.94 &  1.09 &  0.00 & 30.92 &  6.65 &  0.00 &  0.00 &  0.12 &  1.43 &  0.39 & 27.15 &  0.00 & 14.21 &  0.57 &  1.72 & 15.69 &  0.06 \\
CJS & 2 & 1.01 & 0.17 &  1.09 &  0.00 & 31.50 &  6.13 &  0.00 &  0.00 &  0.12 &  1.43 &  0.39 & 27.11 &  0.00 & 14.19 &  0.59 &  1.72 & 15.67 &  0.06 \\
CJS & 2 & 1.57 & 1.29 &  1.20 &  0.00 & 31.40 &  7.31 &  0.00 &  0.00 &  0.12 &  1.54 &  0.38 & 26.50 &  0.00 & 13.87 &  0.57 &  1.68 & 15.36 &  0.07 \\
\end{longtable}
\end{landscape}

\end{document}